\documentclass[prodmode,acmcsur]{acmsmall}

\usepackage{ifpdf}
\usepackage{times}
\usepackage{fancyhdr}
\usepackage[ruled]{algorithm2e}

\SetAlFnt{\small}
\SetAlCapFnt{\small}
\SetAlCapNameFnt{\small}
\SetAlCapHSkip{0pt}
\IncMargin{-\parindent}
\usepackage{paralist}
\usepackage{graphicx,epsfig} % more modern
\usepackage{hyperref}    % Better email address and url handling
\usepackage{amsmath,amssymb,bm} % amsthm
\usepackage{multirow}

\acmVolume{V}
\acmNumber{N}
\acmArticle{A}
\acmYear{YEAR}
\acmMonth{12}

\begin{document}

\markboth{Li and Hoi}{Online Portfolio Selection: A Survey}

\title{Online Portfolio Selection: A Survey}

\author{BIN LI and STEVEN C. H. HOI\affil{Nanyang Technological University, Singapore}
%\and%VIVEKANAND GOPALKRISHNAN\affil{Deloitte Analytics Institute}
}
%\author{BIN LI\affil{Wuhan University, P.R. China}
%\and STEVEN C. H. HOI\affil{Nanyang Technological University, Singapore}
%}

\begin{abstract}
Online portfolio selection is a fundamental problem in computational finance, which has been extensively studied across several research communities, including finance, statistics, artificial intelligence, machine learning, and data mining, etc. This article aims to provide a comprehensive survey and a structural understanding of published online portfolio selection techniques. From an online machine learning perspective, we first formulate online portfolio selection as a sequential decision problem, and then survey a variety of state-of-the-art approaches, which are grouped into several major categories, including benchmarks, ``Follow-the-Winner" approaches, ``Follow-the-Loser" approaches, ``Pattern-Matching" based approaches, and ``Meta-Learning Algorithms". In addition to the problem formulation and related algorithms, we also discuss the relationship of these algorithms with the Capital Growth theory in order to better understand the similarities and differences of their underlying trading ideas. This article aims to provide a timely and comprehensive survey for both machine learning and data mining researchers in academia and quantitative portfolio managers in the financial industry to help them understand the state-of-the-art and facilitate their research and practical applications. We also discuss some open issues and evaluate some emerging new trends for future research directions.
\end{abstract}




\acmformat{Li, B., Hoi, S.~C.H., 2012. OnLine Portfolio Selection: A Survey.}

\begin{bottomstuff}
This work is fully supported by Singapore MOE Academic tier-1 grant (RG33/11).

More information about the OLPS project is available at:~\url{http://OLPS.stevenhoi.org/}.

Author's address: Li, B. {and} Hoi, S.~C.H., School of Computer Engineering, Nanyang Technological University,  Singapore.
  % Gopalkrishnan, V., Deloitte Analytics Institute (Asia), Singapore.\\
E-mail: \{binli, chhoi\}@ntu.edu.sg.
%;vivek@deloitte.com.
\end{bottomstuff}

\maketitle

% --------------------------------------------------------------
\section{Introduction}
\label{sec:introduction}

Portfolio selection, aiming to optimize the allocation of wealth across a set of assets, is a fundamental research problem in computational finance and a practical engineering task in financial engineering. There are two major schools for investigating this problem, that is, the Mean Variance Theory~\cite{Markowitz52,Markowitz59,MVA00} mainly from the finance community and Capital Growth Theory~\cite{Kelly56,HZ95} primarily originated from information theory. The Mean Variance Theory, widely known in asset management industry, focuses on a single-period (batch) portfolio selection to trade off a portfolio's expected return (mean) and risk (variance), which typically determines the optimal portfolios subject to the investor's risk-return profile. On the other hand,  Capital Growth Theory focuses on multiple-period or sequential portfolio selection, aiming to maximize the portfolio's expected growth rate, or expected $\log$ return. While both theories solve the task of portfolio selection, the latter is fitted to the ``online" scenario, which naturally consists of multiple periods and is the focus of this article.

Online portfolio selection, which sequentially selects a portfolio over a set of assets in order to achieve certain targets, is a natural and important task for asset portfolio management. Aiming to maximize the cumulative wealth, several categories of algorithms have been proposed to solve this task. One category of algorithms, termed ``Follow-the-Winner", tries to asymptotically achieve the same growth rate (expected $\log$ return) as that of an optimal strategy, which is often based on the Capital Growth Theory. The second category, named ``Follow-the-Loser", transfers the wealth from winning assets to losers, which seems contradictory to the common sense but empirically often achieves significantly better performance. Finally, the third category, termed ``Pattern-Matching" based approach, tries to predict the next market distribution based on a sample of historical data and explicitly optimizes the portfolio based on the sampled distribution. While the above three categories are focused on a single strategy (class), there are also some other strategies that focus on combining multiple strategies (classes), termed as ``Meta-Learning Algorithms". As a brief summary, Table~\ref{tab:algorithms_classification} outlines the list of main algorithms and corresponding references.

This article provides a comprehensive survey of online portfolio selection algorithms belonging to the above categories.  To the best of our knowledge, this is the first survey that includes the above three categories and the meta-learning algorithms as well. Moreover, we are the first to explicitly discuss the connection between the online portfolio selection algorithms and Capital Growth Theory, and illustrate their underlying trading ideas. In the following sections, we also clarify the scope of this article and discuss some related existing surveys in the literature.

\begin{table*}
\tbl{General classification for the state-of-the-art online portfolio selection  algorithms.\label{tab:algorithms_classification}}{
  \begin{tabular}{l|l|l}
    \hline
    Classifications & Algorithms   & Representative References \\
    \hline
    \multirow{3}{*}{Benchmarks} & Buy And Hold & \\
        & Best Stock & \\
        & Constant Rebalanced Portfolios & \citeN{Kelly56};~\citeN{Cover91}\\
     \hline
    \multirow{5}{*}{Follow-the-Winner}     & Universal Portfolios & \citeN{Cover91};~\citeN{CO96}\\
    & Exponential Gradient & \citeANP{HSS+96}\citeyear{HSS+96,HSS+98}\\
    & Follow the Leader & \citeN{GS00}\\
    & Follow the Regularized Leader & \citeN{AHK+06} \\
    & Aggregating-type Algorithms & \citeN{VW98}\\
    \hline
    \multirow{5}{*}{Follow-the-Loser}     & Anti Correlation  & \citeANP{BEG03}\citeyear{BEG03,BEG04} \\
    & Passive Aggressive Mean Reversion  & \citeN{LZH+12} \\
    & Confidence Weighted Mean Reversion  & \citeANP{LHZ+11b}\citeyear{LHZ+11b,LHZ+13} \\
    & Online Moving Average Reversion  & \citeN{LH12} \\
    & Robust Median Reversion  & \citeN{HZL+13} \\
    \hline
    \multirow{7}{*}{Pattern-Matching Approaches}    &  Nonparametric Histogram Log-optimal  Strategy & \citeN{GLU06}\\
      & Nonparametric Kernel-based Log-optimal Strategy & \citeN{GLU06}\\
    & Nonparametric Nearest Neighbor Log-optimal Strategy & \citeN{GUW08}\\
    & Correlation-driven Nonparametric Learning Strategy & \citeN{LHG11a}\\
    & Nonparametric Kernel-based Semi-log-optimal Strategy & \citeN{GUV07}\\
    & Nonparametric Kernel-based Markowitz-type Strategy & \citeN{OV07}\\
    & Nonparametric Kernel-based GV-type Strategy & \citeN{GV08}\\
    \hline
    \multirow{5}{*}{Meta-Learning Algorithms} & Aggregating Algorithm  & \citeN{Vovk90}\citeyear{VW98} \\
    & Fast Universalization Algorithm & \citeANP{ADK02}\citeyear{ADK02,ADM04} \\
    & Online Gradient Updates & \citeN{DB11} \\
    & Online Newton Updates & \citeN{DB11} \\
    & Follow the Leading History & \citeN{HS09} \\
    \hline
  \end{tabular}}
\end{table*}

\subsection{Scope}
\label{sec:introduction-scope}

In this survey, we focus on discussing the empirical motivating ideas of the online portfolio selection algorithms, while only skimming theoretical aspects (such as competitive analysis by~\citeN{El-Yaniv98} and~\citeN{BEG00} and asymptotical convergence analysis by~\citeN{GOW12}). Moreover, various other related issues and topics are excluded from this survey, as discussed below.%In the following, we discuss some scopes which will not be covered.

First of all, it is important to mention that the ``Portfolio Selection" task in our survey differs from a great body of financial engineering studies~\cite{KAY+93,MF98,CT03,LLC09,Dhar11,HLT11}, which attempted to forecast financial time series by applying machine learning techniques and conduct single stock trading~\cite{KM00,DBLP:conf/alt/KoolenV12}, such as reinforcement learning~\cite{MWL+98,MS01,OLZ02}, neural networks~\cite{KAY+93,DPR+01}, genetic algorithms~\cite{MM96,AK99,MJ11}, decision trees~\cite{TYL04}, and support vector machines~\cite{TC02,CT03,LLC09}, boosting and expert weighting~\cite{Creamer07,CF07,CF10,Creamer12}, etc. The key difference between these existing works and subject area of this survey is that their learning goal is to make explicit predictions of future prices/trends and to trade on a single asset~\cite[Section 6]{BEG00}, while our goal is to directly optimize the allocation among a set of assets.
%without explicit price predictions.

Second, this survey emphasizes the importance of ``online" decision for portfolio selection, meaning that related market information arrives sequentially and the allocation decision must be made immediately. Due to the sequential (online) nature of this task, we mainly focus on the survey of multi-period/sequential portfolio selection work, in which the portfolio is rebalanced to a specified allocation at the end of each trading period~\cite{Cover91}, and the goal typically is to \textit{maximize the expected $\log$ return} over a sequence of trading periods. We note that these work can be connected to the Capital Growth Theory~\cite{Kelly56}, stemmed from the seminal paper of~\citeN{Kelly56} and further developed by~\citeANP{Breiman60}\citeyear{Breiman60,Breiman61}, \citeANP{Hakansson70}\citeyear{Hakansson70,Hakansson71},  \citeANP{Thorp71}\citeyear{Thorp69,Thorp71}, \citeN{BC80}, \citeN{FW81}, \citeN{AC88}, \citeN{BC88}, \citeN{MZB92}, \citeN{MZ99}, \citeN{ZZ07}, \citeN{MTZ10}, etc. It has been successfully applied to gambling~\cite{Thorp62,Thorp69,Thorp97}, sports betting~\cite{HZR81,ZH84,Thorp97,ZH08}, and portfolio investment~\cite{TK67,RT92,Ziemba05}. We thus exclude the studies related to the Mean Variance portfolio theory~\cite{Markowitz52,Markowitz59}, which were typically developed for single-period (batch) portfolio selection (except some extensions~\cite{LN00,DXZ10}).

Finally, this article focuses on surveying the algorithmic aspects and providing a structural understanding of the existing online portfolio selection strategies. To prevent loss of focus, we will not dig into theoretical details. In the literature, there is a large body of related work for the theory~\cite{MTZ11}. Interested researchers can explore the details of the theory from two exhaustive surveys~\cite{Thorp97,MZ08}, and its history from~\citeN{Poundstone05} and~\citeN[Chapter 1]{GOW12}.

\subsection{Related Surveys}
\label{sec:introduction-related_surveys}

There exist several related surveys in this area, but none of them is comprehensive and timely enough for understanding the state-of-the-art of online portfolio selection research. For example, \citeN[Section 5]{El-Yaniv98} and \citeN{BEG00} surveyed the online portfolio selection problem in the framework of competitive analysis. Using our classification in Table~\ref{tab:algorithms_classification}, \citeANP{BEG00} mainly surveyed the benchmarks and two Follow-the-Winner algorithms, that is, Universal Portfolios and Exponential Gradient (refer to the details in Section~\ref{sec:algorithms-follow_winner}). Although the competitive framework is important for the Follow-the-Winner category, both surveys are out-of-date in the sense that they do not include a number of state-of-the-art algorithms afterward. A recent survey by \citeN[Chapter 2]{GOW12} mainly surveyed  Pattern-Matching based approaches, i.e., the third category as shown in Table~\ref{tab:algorithms_classification}, which does not include the other categories in this area and is thus far from complete.

\subsection{Organization}
\label{sec:introduction-organization}

The remainder of this article is organized as follows. Section~\ref{sec:problem_setting} formulates the problem of online portfolio selection formally and addresses several practical issues.
Section~\ref{sec:algorithms} introduces the state-of-the-art algorithms, including Benchmarks in Section~\ref{sec:algorithms-benchmarks}, the Follow-the-Winner approaches in Section~\ref{sec:algorithms-follow_winner}, Follow-the-Loser approaches in Section~\ref{sec:algorithms-follow_loser}, Pattern-Matching based Approaches in Section~\ref{sec:algorithms-pattern_based}, and Meta-Learning Algorithms in Section~\ref{sec:algorithms-meta_algorithms}, etc. Section~\ref{sec:cgt} connects the existing algorithms with the Capital Growth Theory and also illustrates the essentials of their underlying trading ideas. Section~\ref{sec:open_issues} discusses several related open issues, and finally Section~\ref{sec:conclusion} concludes this survey and outlines some future directions.

% --------------------------------------------------------------
\section{Problem Setting}
\label{sec:problem_setting}

Consider a financial market with $m$ assets, we invest our wealth over all the assets in the market for a sequence of $n$ trading periods. The market price change is represented by a $m$-dimensional \textit{price relative vector} $\mathbf{x}_{t}\in\mathbb{R}_{+}^{m}, t = 1, \dots, n$, where the $i^{th}$ element of $t^{th}$ price relative vector, $x_{t, i}$, denotes the ratio of $t^{th}$ closing price to last closing price for the $i^{th}$ assets. Thus, an investment in asset $i$ on period $t$ increases by a factor of $x_{t, i}$. We also denote the market price changes from period $t_{1}$ to $t_{2}$ ($t_{2}>t_{1}$) by a \textit{market window}, which consists of a sequence of price relative vectors $\mathbf{x}_{t_{1}}^{t_{2}}=\left\{ \mathbf{x}_{t_{1}}, \dots, \mathbf{x}_{t_{2}}\right\}$, where $t_{1}$ denotes the beginning period and $t_{2}$ denotes the ending period. One special market window starts from period $1$ to $n$, that is, $\mathbf{x}_{1}^{n}=\left\{ \mathbf{x}_{1}, \dots, \mathbf{x}_{n}\right\}$.

At the beginning of the $t^{th}$ period, an investment is specified by a \textit{portfolio vector} $\mathbf{b}_{t}, t = 1, \dots, n$.
The $i^{th}$ element of $t^{th}$ portfolio, $b_{t, i}$, represents the proportion of capital invested in the $i^{th}$ asset.
Typically, we assume a portfolio is self-financed and no margin/short is allowed.
Thus, a portfolio satisfies the constraint that each entry is non-negative and all entries sum up to one, that is, $\mathbf{b}_{t}\in \Delta_{m}$, where $\Delta_{m}=\left\{\mathbf{b}:\mathbf{b}\succeq 0, \mathbf{b}^{\top}\mathbf{1}=1\right\}$.
Here, $\mathbf{1}$ is the $m$-dimensional vector of all $1$s, and $\mathbf{b}^{\top}\mathbf{1}$ denotes the inner product of $\mathbf{b}$ and $\mathbf{1}$.
The investment procedure from period $1$ to $n$ is represented by a \textit{portfolio strategy}, which is a sequence of mappings as follows:
\begin{equation}\nonumber
\mathbf{b}_{1}=\frac{1}{m}\mathbf{1}, \quad
  \mathbf{b}_{t}:\mathbb{R}_{+}^{m\left(t-1\right)}\rightarrow
  \Delta_{m}, \; t =2, 3, \dots, n,
\end{equation}
where ${\mathbf{b}}_{t}={\mathbf{b}}_{t}\left(\mathbf{x}_{1}^{t-1}\right)$
denotes the portfolio computed from the past market window $\mathbf{x}_{1}^{t-1}$.
Let us denote the portfolio strategy for $n$ periods as $\mathbf{b}_{1}^{n}=\left\{ \mathbf{b}_1, \ldots, \mathbf{b}_{n} \right\}$.

For the $t^{th}$ period, a portfolio manager apportions its capital according to portfolio $\mathbf{b}_{t}$ at the opening time, and holds the portfolio until the closing time.
Thus, the portfolio wealth will increase by a factor of $\mathbf{b}_{t}^{\top}\mathbf{x}_{t}=\sum_{i=1}^{m} b_{t, i}x_{t, i}$. Since this model uses price relatives and re-invests the capital, the portfolio wealth will increase multiplicatively. From period $1$ to $n$, a portfolio strategy $\mathbf{b}_1^{n}$ increases the initial wealth ${S}_{0}$ by a factor of $\prod_{t=1}^{n} \mathbf{b}_{t}^{\top} \mathbf{x}_{t}$, that is, the final \textit{cumulative wealth} after a sequence of $n$ periods is
\begin{equation}\nonumber
{S}_{n}\left(\mathbf{b}_{1}^{n}\right) =
{S}_{0} \prod_{t=1}^{n} \mathbf{b}_{t}^{\top} \mathbf{x}_{t} = {S}_{0} \prod_{t=1}^{n} \sum_{i=1}^{m} b_{t, i}x_{t, i}.
\end{equation}
Since the model assumes multi-period investment, we define the \textit{exponential growth rate} for a strategy $\mathbf{b}_{1}^{n}$ as,
\begin{equation}\nonumber
  {W}_{n}\left(\mathbf{b}_{1}^{n}\right) =\frac{1}{n}\log
  {S}_{n}\left(\mathbf{b}_{1}^{n}\right)=\frac{1}{n}\sum_{t=1}^{n}\log \mathbf{b}_{t}\cdot \mathbf{x}_{t}.
\end{equation}

Finally, let us combine all elements and formulate the online portfolio selection model. In a portfolio selection task the decision maker is a portfolio manager, whose goal is to produce a portfolio strategy $\mathbf{b}_{1}^{n}$ in order to achieve certain targets. Following the principle as the algorithms shown in Table~\ref{tab:algorithms_classification}, our target is to maximize the portfolio cumulative wealth ${S}_{n}$. The portfolio manager computes the portfolio strategy in a sequential fashion. On the beginning of period $t$, based on previous market window $\mathbf{x}_{1}^{t-1}$, the portfolio manager learns a new portfolio vector ${\mathbf{b}}_{t}$ for the coming price relative vector ${\mathbf{x}}_{t}$, where the decision criterion varies among different managers/strategies. The portfolio $\mathbf{b}_{t}$ is scored using the portfolio period return $\mathbf{b}_{t}\cdot \mathbf{x}_{t}$. This procedure is repeated until period $n$, and the strategy is finally scored according to the portfolio cumulative wealth ${S}_{n}$. Algorithm~\ref{alg:problem_setting-model} shows the framework of online portfolio selection, which serves as a general procedure to backtest any online portfolio selection algorithm.

\begin{algorithm}[t]
    \SetAlgoLined
    \KwIn{$\mathbf{x}_{1}^{n}$: Historical market sequence}
    \KwOut{${S}_{n}$: Final cumulative wealth}
    \BlankLine
    Initialize  ${{S}}_{0}=1,{\mathbf{b}}_{1}=\left(\frac{1}{m}, \ldots, \frac{1}{m}\right)$ \\
    \For{$t=1, 2, \ldots, n$}{
        Portfolio manager computes a portfolio ${\mathbf{b}}_{t}$ \;
        Market reveals the market price relative ${\mathbf{x}}_{t}$ \;
        Portfolio incurs period return ${\mathbf{b}}_{t}^{\top}
        {\mathbf{x}}_{t}$ and updates cumulative return ${S}_{t}={S}_{t-1}\times\left({\mathbf{b}}_{t}^{\top}
        {\mathbf{x}}_{t}\right)$ \;
        Portfolio manager updates his/her online portfolio selection rules \;
    }
    \caption{Online portfolio selection framework.}
    \label{alg:problem_setting-model}
\end{algorithm}

% 2.1.3. Model constraints
In general, some assumptions are made in the above widely adopted model:
\begin{enumerate}
\item Transaction cost: we assume no transaction costs/taxes in the model;
\item Market liquidity: we assume that one can buy and sell any quantity of any asset in its closing prices;
\item Impact cost: we assume market behavior is not affected by any portfolio selection strategy.
\end{enumerate}

To better understand the notions and model above, let us illustrate with a classical example.
\begin{example}[Synthetic market by~\citeN{CG86}]\label{example:cg86}
Assume a two-asset market with cash and one volatile asset with the price relative sequence $\mathbf{x}_{1}^{n}=\left\{ \left(1, 2\right), \left(1, \frac{1}{2}\right), \left(1, 2\right), \dots  \right\}$.
The $1^{st}$ price relative vector $\mathbf{x}_{1}=\left(1, 2\right)$ means that if we invest $\$1$ in the first asset, you will get $\$1$ at the end of period; if we invest $\$1$ in the second asset, we will get $\$2$ after the period.

Let a fixed proportion portfolio strategy be $\mathbf{b}_{1}^{n}=\left\{ \left(\frac{1}{2}, \frac{1}{2}\right), \left(\frac{1}{2}, \frac{1}{2}\right), \dots \right\}$, which means everyday the manager redistributes the capital equally among the two assets.
%Its first portfolio is $\mathbf{b}_{1}=\left(\frac{1}{2}, \frac{1}{2}\right)$.
For the $1^{st}$ period, the portfolio wealth increases by a factor of $1\times \frac{1}{2}+2\times \frac{1}{2}=\frac{3}{2}$.
Initializing the capital with ${S}_{0}=1$, then the capital at the end of the $1^{st}$ period equals ${S}_{1}={S}_{0}\times \frac{3}{2}=\frac{3}{2}$.
Similarly, ${S}_{2}={S}_{1}\times \left(1\times\frac{1}{2} + \frac{1}{2}\times\frac{1}{2}\right)=\frac{3}{2}\times\frac{3}{4}=\frac{9}{8}$.
Thus, at the end of  period $n$, the final cumulative wealth equals,
\begin{equation}\nonumber
{S}_{n}\left(\mathbf{b}_{1}^{n}\right)=\begin{cases}
\frac{9}{8}^{\frac{n}{2}} &  n\mbox{ is even}\\
\frac{3}{2}\times\frac{9}{8}^{\frac{n-1}{2}} & n\mbox{ is odd}\\
\end{cases},
\end{equation}
and the exponential growth rate is,
\begin{equation}\nonumber
{W}_{n}\left(\mathbf{b}_{1}^{n}\right)=\begin{cases}
\frac{1}{2}\log\frac{9}{8} &  n\mbox{ is even}\\
\frac{n-1}{2n}\log\frac{9}{8}+\frac{1}{n}\log\frac{3}{2} & n\mbox{ is odd}\\
\end{cases},
\end{equation}
which approaches $\frac{1}{2}\log\frac{9}{8} > 0$ if $n$ is sufficiently large.
\end{example}

\subsection{Transaction Cost}
\label{sec:problem_setting-transaction_cost}

% 2.3.1. transaction cost and its model
In reality, the most important and unavoidable issue is \emph{transaction costs}.
In this section, we model the transaction costs into our formulation, which enables us to evaluate an online portfolio selection algorithms.
However, we will not introduce strategies~\cite{DN90,IC00,AST01,Schafer02,GV08,OU11} that directly solve the transaction costs issues.

The widely adopted transaction costs model is the \textit{proportional transaction costs} model~\cite{BK99,GV08}, in which the incurred transaction cost is proportional to the wealth transferred during rebalancing.
Let the brokers charge transaction costs on both buying and selling.
At the beginning of the $t^{th}$ period, the portfolio manager intends to rebalance the portfolio from closing price adjusted portfolio $\hat{\mathbf{b}}_{t-1}$ to a new portfolio $\mathbf{b}_{t}$.
Here $\hat{\mathbf{b}}_{t-1}$ is calculated as,
$ \hat{b}_{t-1, i}=\frac{b_{t-1, i}x_{t-1, i}}{\mathbf{b}_{t-1}\cdot \mathbf{x}_{t-1}}, i = 1, \dots, m$.
Assuming two transaction cost rates $\gamma_{b}\in \left(0, 1\right)$ and $\gamma_{s}\in \left( 0, 1\right)$, where $\gamma_{b}$ denotes the transaction costs rate incurred during buying and $\gamma_{s}$ denotes the transaction costs rate incurred during  selling.
After rebalancing, $S_{t-1}$ will be decomposed into two parts, that is,  the net wealth $N_{t-1}$ in the new portfolio $\mathbf{b}_{t}$ and the transaction costs incurred during the buying and selling.
If the wealth on asset $i$ before rebalancing is higher than that after reblancing, that is, $\frac{b_{t-1, i}x_{t-1,i}}{\mathbf{b}_{t-1}\cdot\mathbf{x}_{t-1}}S_{t-1}\geq b_{t, i}N_{t-1}$, then there will be a selling rebalancing.
Otherwise, then a buying rebalancing is required.
Formally,
\begin{equation}\nonumber
S_{t-1}=N_{t-1} + \gamma_{s}\sum_{i=1}^{m}\left(\frac{b_{t-1, i}x_{t-1,i}}{\mathbf{b}_{t-1}\cdot\mathbf{x}_{t-1}}S_{t-1}-b_{t, i}N_{t-1}\right)^{+} + \gamma_{b} \sum_{i=1}^{m}\left(b_{t, i}N_{t-1}- \frac{b_{t-1, i}x_{t-1,i}}{\mathbf{b}_{t-1}\cdot\mathbf{x}_{t-1}}S_{t-1}  \right)^{+}.
\end{equation}
Let use denote  \textit{transaction costs factor}~\cite{GV08} as the ratio of  net wealth after rebalancing to wealth before rebalancing, that is, $c_{t-1}=\frac{N_{t-1}}{S_{t-1}} \in \left(0, 1\right)$.
Dividing above equation by $S_{t-1}$, we can get,
\begin{equation}\label{eq:problem_setting-transaction_cost}
1 = c_{t-1} + \gamma_{s}\sum_{i=1}^{m}\left(\frac{b_{t-1, i}x_{t-1,i}}{\mathbf{b}_{t-1}\cdot\mathbf{x}_{t-1}}-b_{t, i}c_{t-1}\right)^{+}+\gamma_{b} \sum_{i=1}^{m}\left(b_{t, i}c_{t-1}- \frac{b_{t-1, i}x_{t-1,i}}{\mathbf{b}_{t-1}\cdot\mathbf{x}_{t-1}}  \right)^{+}.
\end{equation}
Clearly, given $\mathbf{b}_{t-1}$,  $\mathbf{x}_{t-1}$, and $\mathbf{b}_{t}$, there exists a unique transaction costs factor for each rebalancing.
Thus, we can denote $c_{t-1}$ as a function, $c_{t-1}=c\left(\mathbf{b}_{t}, \mathbf{b}_{t-1}, \mathbf{x}_{t-1}\right)$.
Moreover, considering the portfolio is in the simplex domain, then the factor ranges between $\frac{1-\gamma_{s}}{1+\gamma_{b}}\leq c_{t-1}\leq 1$.

Finally, for each period $t$, the wealth grows by a factor  as,
\begin{equation}\nonumber
S_{t}=S_{t-1}\times c_{t-1}\times \left(\mathbf{b}_{t}\cdot \mathbf{x}_{t} \right),
\end{equation}
 and the final cumulative wealth after $n$ periods equals,
\begin{equation}\nonumber
S_{n}=S_{0}\prod_{t=1}^{n}c_{t-1}\times \left(\mathbf{b}_{t}\cdot \mathbf{x}_{t} \right),
\end{equation}
where $c_{t-1}$ is calculated as Eq.~\eqref{eq:problem_setting-transaction_cost}.
\section{Online Portfolio Selection Approaches}
\label{sec:algorithms}

In this section, we survey the area of online portfolio selection. Algorithms in this area formulate the online portfolio selection task as in Section~\ref{sec:problem_setting} and derive  explicit portfolio update schemes for each  period.
Basically, the routine  is to implicitly assume various price relative predictions and learn optimal portfolios.

In the subsequent sections, we mainly list the algorithms following Table~\ref{tab:algorithms_classification}.
In particular, we first introduce several benchmark algorithms in Section~\ref{sec:algorithms-benchmarks}.
Then, we introduce the algorithms with explicit update schemes in the subsequent three sections.
We classifies them based on the direction of the weight transfer.
The first approach, \textit{Follow-the-Winner} approach, tries to increase the relative weights of  more successful experts/stocks, often based on their historical performance.
On the contrary, the second approach, \textit{Follow-the-Loser} approach, tries to increase the relative weights of less successful experts/stocks, or transfer the weights from winners to losers.
The third approach, \textit{Pattern-Matching} based approach, tries to build a portfolio based on some sampled similar historical patterns with no explicit weights transfer directions.
After that, we survey  \textit{Meta-Learning Algorithms}, which can be applied to higher level experts equipped with any existing algorithm.

\subsection{Benchmarks}
\label{sec:algorithms-benchmarks}

\subsubsection{Buy And Hold Strategy}
\label{sec:algorithms-benchmarks-bah}

The most common baseline is \textit{Buy-And-Hold} (BAH) strategy, that is, one invests wealth among a pool of assets with an initial portfolio ${\mathbf{b}}_{1}$ and holds the portfolio until the end.
The manager only buys the assets at the beginning of the $1^{st}$ period and does not rebalance in the following periods, while the portfolio holdings are implicitly changed following the market fluctuations.
For example, at the end of the $1^{st}$ period, the portfolio holding becomes $\frac{\mathbf{b}_{1}\bigodot\mathbf{x}_{1}}{\mathbf{b}_{1}^{\top}\mathbf{x}_{1}}$, where $\bigodot$ denotes element-wise product.
In a summary, the final cumulative wealth achieved by a BAH strategy is initial portfolio weighted average of individual stocks' final wealth,
\begin{equation}\nonumber
{S}_{n}\left(BAH\left(\mathbf{b}_{1}\right)\right)= \mathbf{b}_{1}\cdot \left(\bigodot_{t=1}^{n}\mathbf{x}_{t}\right).
\end{equation}

The BAH strategy with  initial \textit{uniform} portfolio ${\mathbf{b}}_{1}=\left(\frac{1}{m}, \dots, \frac{1}{m}\right)$ is referred to as \textit{uniform BAH} strategy, which is often adopted as  a \textit{market} strategy to produce a market index.

\subsubsection{Best Stock Strategy}
\label{sec:algorithms-benchmarks-best}

Another widely adopted benchmark is the \textit{Best Stock} (Best) strategy, which is a special BAH strategy that puts all capital on the stock with best performance in hindsight.
Clearly, its initial portfolio $\mathbf{b}^{\circ}$  in hindsight can be calculated as,
\begin{equation}\nonumber
\mathbf{b}^{\circ} = \mathop{\arg\max}_{\mathbf{b}\in \Delta_{m}}\mathbf{b} \cdot \left(\bigodot_{t=1}^{n}\mathbf{x}_{t}\right).
\end{equation}
As a result, the final cumulative wealth achieved by the Best strategy can be calculated as,
\begin{equation}\nonumber
\begin{split}
{S}_{n}\left(Best\right) &= \max_{\mathbf{b}\in\Delta_{m}}\mathbf{b} \cdot \left(\bigodot_{t=1}^{n}\mathbf{x}_{t}\right) ={S}_{n}\left(BAH\left(\mathbf{b}^{\circ}\right)\right).\\
\end{split}
\end{equation}

\subsubsection{Constant Rebalanced Portfolios}
\label{sec:algorithms-benchmarks-crp}

Another more challenging benchmark strategy is the \textit{Constant Rebalanced Portfolio} (CRP) strategy, which rebalances the portfolio to a fixed portfolio $\mathbf{b}$ every period.
In particular, the portfolio strategy can be represented as $\mathbf{b}_{1}^{n}=\left\{\mathbf{b}, \mathbf{b}, \dots \right\}$.
Thus, the cumulative portfolio wealth achieved by a CRP strategy after $n$  periods is defined as,
\begin{equation}\nonumber
{S}_{n}\left(CRP\left(\mathbf{b}\right) \right) =
\prod_{t=1}^{n} \mathbf{b}^{\top}\mathbf{x}_{t}.
\end{equation}
One special CRP strategy that rebalances to uniform portfolio $\mathbf{b}=\left(\frac{1}{m}, \dots, \frac{1}{m}\right)$ each period is named \textit{Uniform Constant Rebalanced Portfolios} (UCRP).
It is possible to calculate an optimal offline portfolio for the CRP strategy as,
\begin{equation}\nonumber
\mathbf{b}^{\star}=\mathop{\arg \max}_{\mathbf{b}^{n}\in\Delta_{m}}
\log {S}_{n}\left(CRP\left(\mathbf{b}\right)\right)=\mathop{\arg \max}_{\mathbf{b}\in\Delta_{m}}
\sum_{t=1}^{n}\log\left(\mathbf{b}^{\top}\mathbf{x}_{t}\right),
\end{equation}
which is convex and can be efficiently solved.
The CRP strategy with $\mathbf{b}^{\star}$ is denoted by \textit{Best Constant Rebalanced Portfolio} (BCRP).
BCRP achieves a final cumulative portfolio wealth and corresponding exponential growth rate defined as follows,
\begin{equation}\nonumber
\begin{split}
{S}_{n}\left(BCRP \right) &= \max_{\mathbf{b}\in\Delta_{m}} {S}_{n}\left(CRP\left(\mathbf{b}\right) \right) ={S}_{n}\left(CRP\left(\mathbf{b}^{\star}\right)\right),\\
{W}_{n}\left(BCRP\right) &=\frac{1}{n}\log {S}_{n}\left(BCRP\right)=\frac{1}{n}\log
{S}_{n}\left(CRP\left(\mathbf{b}^{\star}\right)\right).
\end{split}
\end{equation}

Note that BCRP strategy is a hindsight strategy, which can only be calculated with complete market sequences.
\citeN{Cover91} proved the benefits of BCRP as a target, that is, BCRP exceeds the best stock, Value Line Index (geometric mean of component returns) and the Dow Jones Index (arithmetic mean of component returns, or BAH).
Moreover, BCRP is invariant under permutations of the price relative sequences, i.e., it does not depend on the order in which $\mathbf{x}_{1}, \mathbf{x}_{2}, \dots, \mathbf{x}_{n}$ occur.

Till now let us compare BAH and CRP strategy by continuing the Example~\ref{example:cg86}.

\begin{example}[Synthetic market by~\citeN{CG86}]\label{example:bcrp}
Assume a two-asset market with cash and one volatile asset with the price relative sequence $\mathbf{x}_{1}^{n}=\left\{ \left(1, 2\right), \left(1, \frac{1}{2}\right), \left(1, 2\right), \dots  \right\}$.
Let us consider BAH with uniform initial portfolio $\mathbf{b}_{1}=\left( \frac{1}{2}, \frac{1}{2} \right)$ and the CRP with uniform portfolio $\mathbf{b}=\left( \frac{1}{2}, \frac{1}{2}\right)$.
Clearly, since no asset grows in the long run, the final wealth of BAH equals the uniform weighted summation of two assets, which roughly equals to $1$ in the long run.
On the other hand, according to the analysis of Example~\ref{example:cg86}, the final cumulative wealth of CRP is roughly $\frac{9}{8}^{\frac{n}{2}}$, which increases exponentially.
Note that the BAH only rebalances on the $1^{st}$ period, while the CRP rebalances every period.
On the same synthetic market, while market provides no return and CRP can produce an exponentially increasing return.
The underlying idea of CRP is to take advantage of the underlying volatility, or so-called ``volatility pumping"~\cite[Chapter 15]{Luenberger98}.
\end{example}

Since CRP rebalances a fixed portfolio each period, its frequent transactions will incur high transaction costs.
\citeANP{HSS+96}~\citeyear{HSS+96,HSS+98} proposed a \textit{Semi-Constant Rebalanced Portfolio} (Semi-CRP), which rebalances the portfolio on selected periods rather than every period.

One desired theoretical result for online portfolio selection is ``\textit{universality}''~\cite{Cover91}.
An online portfolio selection algorithm $Alg$ is \emph{universal} if its average
  \textit{(external) regret}~\cite{SL05,BM07} for $n$  periods asymptotically approaches $0$,
\begin{equation}\label{eq:algorithms-benchmarks-regret}
    \frac{1}{n}\mbox{regret}_{n}\left(Alg\right)={W}_{n}\left(BCRP\right)-{W}_{n}\left(Alg\right)
  \longrightarrow 0, \mbox{ as } n\rightarrow \infty.
\end{equation}
In other words, a {universal} portfolio selection algorithm asymptotically approaches the same exponential growth rate as BCRP strategy for arbitrary sequences of price relatives.

\subsection{Follow-the-Winner Approaches}
\label{sec:algorithms-follow_winner}

The first approach, \textit{Follow-the-Winner}, is characterized by increasing the relative weights of more successful experts/stocks.
Rather than targeting market and best stock, algorithms in this category often aim to track the BCRP strategy, which can be shown to be the optimal strategy in an i.i.d. market~\cite[Theorem 15.3.1]{CT91}.
On other words, such optimality motivates that ``universal portfolio selection" algorithms approach the performance of the hindsight BCRP for arbitrary sequence of  price relative vectors, called individual sequences.
%On other words, an algorithm in this category aims to be universal portfolio selection algorithm.

\subsubsection{Universal Portfolios}
\label{sec:algorithms-follow_winner-up}

The basic idea of \textit{Universal Portfolio}-type algorithms is to assign the capital to a single class of base experts, let the experts run, and finally pool their wealth.
Strategies in this type are analogous to the \textit{Buy And Hold} (BAH) strategy.
Their difference is that base BAH expert is the strategy investing on a single stock and thus  the number of experts is the same as that of stocks.
In other words, BAH strategy buys the individual stocks and lets the stocks go and finally pools their individual wealth.
On the other hand, the base expert in the Follow-the-Winner category can be any strategy class that invests in any set of stocks in the market.
Besides, algorithms in this category are also similar to the \textit{Meta-Learning Algorithms} (MLA) further described in Section~\ref{sec:algorithms-meta_algorithms}, while MLA generally applies to experts of multiple classes.

\citeN{Cover91} proposed the \textit{Universal Portfolio} (UP) strategy and~\citeN{CO96} further refined the algorithm as \textit{$\mu$-Weighted Universal Portfolio}, in which $\mu$ denotes a given distribution on the space of valid portfolio $\Delta_m$.
Intuitively, Cover's UP operates similar to a Fund of Funds (FOF), and its main idea is to buy and hold parameterized CRP strategies over the whole simplex domain.
In particular, it initially invests a proportion of wealth ${\emph{d}\mu\left(\mathbf{b}\right)}$ to each portfolio manager operating CRP strategy with $\mathbf{b}\in\Delta_{m}$, and lets the CRP managers run.
Then, at the end, each manager will grow his wealth to ${S}_{n}\left(\mathbf{b}\right)d\mu\left(\mathbf{b} \right)$.
Finally, Cover's UP pools the individual experts' wealth over the continuum of portfolio strategies.%,  leading to a terminal wealth of $S_{n}\left(UP\right)$ in Eq.~\eqref{eq:algorithms-follow_winner-up_wealth}.
Note that ${S}_{n}\left(\mathbf{b}\right)=\mbox{e}^{n{W}_{n}\left(\mathbf{b}\right)}$, which means that the portfolio grows at an exponential rate of ${W}_{n}\left(\mathbf{b}\right)$.

%Ordentlich10 May be cite.

Formally, its update scheme~\cite[Definition~1]{CO96} can be interpreted as a historical performance weighed average of all valid constant rebalanced portfolios,
\begin{equation}\nonumber
{\mathbf{b}}_{t+1}=\frac{\int_{\Delta_{m}} \mathbf{b} {S}_{t}\left(\mathbf{b}\right) \emph{d}\mu\left(\mathbf{b}\right)}{\int_{\Delta_{m}} {S}_{t}\left(\mathbf{b}\right) \emph{d}\mu\left(\mathbf{b}\right)}.
\end{equation}
Note that at the beginning of period $t+1$, one CRP manager's wealth (historical performance) equals to $S_{t}\left(\mathbf{b}\right)d\mu\left(\mathbf{b}\right)$.
Incorporating the initial wealth of $S_{0}=1$, the final cumulative wealth is weighted average of CRP managers' wealth~\cite[Eq.~(24)]{CO96},
\begin{equation}\label{eq:algorithms-follow_winner-up_wealth}
{S}_{n}\left(UP\right)={\int_{\Delta_{m}} {S}_{n}\left(\mathbf{b}\right) \emph{d} \mu\left(\mathbf{b}\right)}.
\end{equation}
One special case is that $\mu$ equals a uniform distribution, the portfolio update reduces to Cover's UP~\cite[Eq.~(1.3)]{Cover91}.
Another special cases is Dirichlet $\left(\frac{1}{2}, \dots, \frac{1}{2}\right)$ weighted Universal Portfolios~\cite{CO96}, which is proved to be a more optimal allocation.
Alternatively, if a loss function is defined as the negative logarithmic function of portfolio return, Cover's UP  is actually an exponentially weighted average forecaster~\cite{CL06}.

\citeN{Cover91} showed that with suitable smoothness conditions, the average of exponentials grows at the same exponential rate as the maximum, one can asymptotically approach BCRP's exponential growth rate.
The regret achieved by Cover's UP is ${O}\left(m \log n\right)$, and its time complexity is ${O}\left(n^{m}\right)$, where $m$ denotes the number of stocks and $n$ refers to the number of periods.
\citeN{CO96} proved that the $\left(\frac{1}{2}, \dots, \frac{1}{2}\right)$ weighted Universal Portfolios has the same scale of regret bound, but a better constant term~\cite[Theorem 2]{CO96}.

As Cover's UP is based on an ideal market model, one research topic with respect to Cover's UP is to extend the algorithm with various realistic assumptions.
\citeN{CO96} extended the model to include side information, which can be instantiated experts' opinions, fundamental data, etc.
%\citeN{CO98} applied the algorithm with short selling and margin, and~
\citeN{BK99} took account of transaction costs for online portfolio selection and proposed a universal portfolio algorithm to handle the costs.
%Besides, they also proposed Semi Constant Rebalanced Portfolios

Another research topic is to generalize Cover's UP with different underlying base expert classes, rather than the CRP strategy.
\citeN{Jamshidian92} generalized the algorithm for continuous time market and derived the long-term performance of Cover's UP in this setting.
\citeN{VW98} applied \textit{aggregating algorithm} (AA)~\cite{Vovk90} to a finite number of arbitrary investment strategies.
Cover's UP becomes a specialized case of AA when applied to an infinite number of CRPs.
We will further investigate AA in Section~\ref{sec:algorithms-follow_winner-aa}.
%\citeN{OC98} analyzed the minimal ratio of the final wealth achieved by any non-anticipating investment strategy to that of  BCRP strategy and provided a strategy to achieve such optimal ratio.
\citeN{OC98} derived the lower bound of the final wealth achieved by any non-anticipating investment strategy to that of BCRP strategy.
\citeN{CB03} generalized Cover's UP from CRP strategy class to any parameterized target class and proposed a universal strategy that costs a polynomial time.
\citeANP{ADK02}~\citeyear{ADK02,ADM04} extended Cover's UP from the parameterized CRP class to a wide class of investment strategies, including trading strategies operating on a single stock and portfolio strategies operating on the whole stock market.
\citeN{KS11} proposed a similar universal algorithm based on the class of semi-constant rebalanced portfolios~\cite{HSS+96,HSS+98}, which provides good performance with transaction costs.

Rather than intuitive analysis, several work has also been proposed to discuss the connection between Cover's UP with universal prediction~\cite{FMG92}, data compression~\cite{Rissanen83} and Markowitz's mean-variance theory~\cite{Markowitz52,Markowitz59}.
\citeN{Algoet92} discussed the universal schemes for prediction, gambling and portfolio selection.
\citeN{Cover96} and~\citeN{Ordentlich96} discussed the connection of universal portfolio selection and data compression.
\citeN{Belentepe05} presented a statistical view of Cover's UP strategy and connect it with traditional Markowitz's mean-variance portfolio theory~\cite{Markowitz52}.
The authors showed that by allowing short selling and leverage, UP is approximately equivalent to sequential mean-variance optimization; otherwise the strategy is approximately equivalent to constrained sequential optimization.
Though its update scheme is distributional free, UP implicitly estimates the multivariate mean and covariance matrix.

Although Cover's UP has a good theoretical performance guarantee, its implementation costs exponential time in the number of assets, which restricts its practical capability.
To overcome this computational bottleneck,~\citeN{KV02} presented an efficient implementation based on non-uniform random walks that are rapidly mixing.
Their implementation requires a poly running time of $O\left(m^{7} n^{8}\right)$, which is a substantial improvement of the original bound of $O\left(n^{m}\right)$.

% Wednesday, March 28, 2012 5:30 pm
\subsubsection{Exponential Gradient}
\label{sec:algorithms-follow_winner-eg}

The strategies in the \textit{Exponential Gradient}-type generally focus on the following optimization problem,
\begin{equation}\label{eq:algorithms-benchmarks-eg}
\mathbf{b}_{t+1}=\mathop{\arg\max}_{\mathbf{b}\in\Delta_{m}}\quad \eta\log \mathbf{b}\cdot \mathbf{x}_{t} - R\left(\mathbf{b}, \mathbf{b}_{t}\right),
\end{equation}
where $R\left(\mathbf{b}, \mathbf{b}_{t}\right)$ denotes a regularization term and $\eta>0$ is the learning rate.
One straightforward interpretation of the optimization is to track the stock with the best performance in last period but keep the new portfolio close to the previous portfolio.
This is obtained using the regularization term $R\left(\mathbf{b}, \mathbf{b}_{t}\right)$.
%and keep the previous portfolio information via a regularization term.

\citeANP{HSS+96}\citeyear{HSS+96,HSS+98} proposed the \textit{Exponential Gradient} (EG) strategy, which is based on the algorithm proposed for  mixture estimation problem~\cite{HSS+97}.
%The EG strategy adopts relative entropy as the regularization term, that is,
The EG strategy employs the relative entropy as the regularization term in Eq.~\eqref{eq:algorithms-benchmarks-eg},
\begin{equation}\nonumber
R\left(\mathbf{b}, \mathbf{b}_{t}\right)=\sum_{i=1}^{m}b_{i}\log\frac{b_{i}}{b_{t, i}}.
\end{equation}
EG's formulation is thus convex in $\mathbf{b}$, however, it is hard to solve since $\log$ is nonlinear.
Thus, the authors adopted $\log$'s first-order Taylor expansion at $\mathbf{b}_{t}$,
\begin{equation}\nonumber
\log\mathbf{b}\cdot\mathbf{x}_{t}\approx\log\left(\mathbf{b}_{t}\cdot \mathbf{x}_{t}\right)+\frac{\mathbf{x}_{t}}{\mathbf{b}_{t}\cdot \mathbf{x}_{t}}\left(\mathbf{b}-\mathbf{b}_{t}\right),
\end{equation}
with which the first term in Eq.~\eqref{eq:algorithms-benchmarks-eg} becomes linear and easy to solve.
Solving the optimization, the update rule~\cite[Eq.~(3.3)]{HSS+98} becomes,
\begin{equation}\nonumber
 {b}_{t+1, i}={{b}_{t, i}\exp \left( \eta
    \frac{{x}_{t, i}}{\mathbf{b}_{t}\cdot
      \mathbf{x}_{t}}\right)}/{Z},\quad i = 1, \dots, m,
\end{equation}
where $Z$ denotes the normalization term such that the portfolio sums to $1$.

The optimization problem~\eqref{eq:algorithms-benchmarks-eg} can also be solved using the \textit{Gradient Projection} (GP) and \textit{Expectation Maximization} (EM) method~\cite{HSS+97}.
GP and EM adopt different regularization terms. In particular, GP adopts L2-norm regularization, and EM adopts $\chi^{2}$ regularization,
\begin{equation}\nonumber
R\left( \mathbf{b}, \mathbf{b}_{t}\right)=\begin{cases}
\frac{1}{2}\sum_{i=1}^{m}\left(b_{i}-b_{t, i}\right)^{2} & \mbox{GP}\\
\frac{1}{2}\sum_{i=1}^{m}\frac{\left(b_{i}-b_{t, i}\right)^{2}}{b_{t, i}} & \mbox{EM}\\
\end{cases}.
\end{equation}
The final update rule of GP~\cite[Eq.~(5)]{HSS+97} is,
\begin{equation}\nonumber
b_{t+1, i}=b_{t, i}+\eta\left(\frac{x_{t, i}}{\mathbf{b}_{t}\cdot \mathbf{x}_{t}}-\frac{1}{m}\sum_{i=1}^{m}\frac{x_{t, i}}{\mathbf{b}_{t}\cdot \mathbf{x}_{t}} \right),
\end{equation}
and the update rule of EM~\cite[Eq.~(7)]{HSS+97} is,
\begin{equation}\nonumber
b_{t+1, i}=b_{t, i}\left(\eta \left(\frac{x_{t, i}}{\mathbf{b}_{t}\cdot \mathbf{x}_{t}}-1\right)+1\right),
\end{equation}
which can also be viewed as the first order approximation of EG's update formula.

The regret of the EG strategy can be bounded by ${O}\left(\sqrt{n \log m}\right)$ with ${O}\left(m\right)$ running time per period.
The regret is not as tight as that of Cover's UP, however, its linear running time substantially surpasses that of Cover's UP.
Besides, the authors also proposed a variant, which has a regret bound of ${O}\left(m^{0.5}(\log m)^{0.25} n^{0.75}\right)$.
Though not proposed for online portfolio selection task, according to~\citeN{HSS+97}, GP can straightforwardly achieve a regret of ${O}\left(\sqrt{mn}\right)$, which is significantly worse than that of EG.

One key parameter for EG is the learning rate $\eta>0$.  In order to achieve the desired regret bound above, $\eta$ has to be small.  However, as $\eta\rightarrow 0$, its update approaches uniform portfolio, and EG reduces to UCRP.

%To consider the transaction costs incurred during the rebalance,~\citeN{Lao02} proposed an additive algorithm named ``Additive Return Transaction Costs" (ARTC), which adopts L2-norm to constraint the deviation from last closing price adjusted price relative.

\citeN{DB11} extended the EG algorithm to the sense of meta-learning algorithm named \textit{Online Gradient Updates} (OGU), which will be introduced in Section~\ref{sec:algorithms-meta_algorithms-ogu_onu}.
OGU combines underlying experts such that the overall system can achieve the performance that is no worse than any convex combination of base experts.
%We will introduce OGU in Section~\ref{sec:algorithms-meta_algorithms-ogu_onu}.

\subsubsection{Follow the Leader}
\label{sec:algorithms-follow_winner-ftl}

Strategies in the \textit{Follow the Leader} (FTL) approach try to track the Best Constant Rebalanced Portfolio (BCRP) until time $t$, that is,
\begin{equation}\label{eq:algorithms-follow_winner-ftl}
  \mathbf{b}_{t+1} = \mathbf{b}_{t}^{\ast} = \mathop{\arg\max}_{\mathbf{b}\in \Delta_{m}} \sum_{j=1}^{t}\log
  \left(\mathbf{b}\cdot \mathbf{x}_{j} \right).
\end{equation}
Clearly, this category follows the BCRP leader, and the ultimate leader is the BCRP over all periods.

\citeN[Chapter 4.4]{Ordentlich96} briefly mentioned a strategy to obtain portfolios by mixing the BCRP up to time $t$ and uniform portfolio,
\begin{equation}\nonumber
  \mathbf{b}_{t+1} = \frac{t}{t+1}\mathbf{b}_{t}^{\ast}+\frac{1}{t+1}\frac{1}{m}\mathbf{1}.
\end{equation}
He also showed its worst case bound, which is slightly worse than that of Cover's UP.

\citeN{GS00} proposed \textit{Successive Constant Rebalanced
  Portfolios} (SCRP) and \textit{Weighted Successive Constant
  Rebalanced Portfolios} (WSCRP) for stationary markets.
For each period, SCRP directly adopts the BCRP portfolio until now, that is,
\begin{equation}\nonumber
  \mathbf{b}_{t+1}= \mathbf{b}_{t}^{\ast}.
\end{equation}
The authors further solved the optimal portfolio $\mathbf{b}_{t}^{\ast}$ via stochastic optimization~\cite{BL97}, resulting in the detail updates of SCRP~\cite[Algorithm 1]{GS00}.
On the other hand, WSCRP outputs a convex combination of SCRP portfolio and last portfolio,
\begin{equation}\nonumber
  \mathbf{b}_{t+1}= \left(1-\gamma\right)\mathbf{b}_{t}^{\ast} + \gamma \mathbf{b}_{t},
\end{equation}
where $\gamma\in\left[0, 1\right]$ represents the trade-off parameter.

The regret bounds achieved by SCRP~\cite[Theorem 1]{GS00} and WSCRP~\cite[Theorem 4]{GS00} are both $O\left(K^{2}\log n\right)$, where $K$ is a uniform upper bound of the gradient of $\log \mathbf{b}^{\top}\mathbf{x}$ with respect to $\mathbf{b}$.
It is straightforward to see that given the same assumption of upper/lower bound of price relatives as Cover's UP~\cite[Theorem 6.1]{Cover91}, the regret bound is on the same scale of Cover's UP, although the constant term is slightly worse.

Rather than assuming that  historical market is stationary, some algorithms assume that the historical market is non-stationary.
\citeN{GS00} propose \textit{Variable Rebalanced Portfolios} (VRP), which calculates the BCRP portfolio based on a latest sliding window. To be more specific, VRP updates its portfolio as follows,
\begin{equation}\nonumber
  \mathbf{b}_{t+1}= \mathop{\arg\max}_{\mathbf{b}\in \Delta_{m}}
  \sum_{j=t-W+1}^{t}\log\left(\mathbf{b}\cdot \mathbf{x}_{j} \right),
\end{equation}
where $W$ denotes a specified window size.
Following their algorithms for Constant Rebalanced Portfolios (CRP), they
further proposed \textit{Successive Variable Rebalanced Portfolios}
(SVRP) and \textit{Weighted Successive Variable Rebalanced Portfolios}
(WSVRP).
No theoretical results were given on the two algorithms.

%SCRP and WSCRP are intuitively reasonable and seem to give good results according to their presentation. However, in the experiments of WSCRP, the authors set high
%values of $\gamma$ ($\gamma = 0.99995, 0.99990, \mbox{ and }0.99950$),
%thus WSCRP almost degrades to UCRP.
%Moreover, though the authors does not present the results of SCRP ($\gamma = 0$), we find that
%SCRP performs not good on the Cover's two-asset portfolios, and even
%bad on Cover's full portfolios compared with UP and EG algorithms.

\citeN{GS03a} further generalized~\citeN{GS00} and proposed
\textit{Adaptive Portfolio Selection} (APS) for online portfolio
selection task. By changing the objective part, APS can handle three types
of portfolio selection task, that is, adaptive Markowitz
portfolio, log-optimal constant rebalanced portfolio, and index
tracking. To handle the transaction cost issue, they proposed
\textit{Threshold Portfolio Selection} (TPS), which only rebalances the
portfolio if the expected return of new portfolio exceeds that of
previous portfolio for more than a threshold.

\subsubsection{Follow the Regularized Leader}
\label{sec:algorithms-follow_winner-ftrl}

Another category of approaches follows the similar idea as FTL, while adding a regularization term, thus actually becomes \textit{Follow the Regularized Leader} (FTRL) approach.
In generally,  FTRL approaches can be formulated as follows,
\begin{equation}\label{eq:algorithms-follow_winner-ftrl}
  \mathbf{b}_{t+1}=\mathop{\arg\max}_{\mathbf{b}\in \Delta_{m}} \sum_{\tau =
    1}^{t} \log \left( \mathbf{b}\cdot \mathbf{x}_{\tau} \right) -
  \frac{\beta}{2}R\left(\mathbf{b}\right),
\end{equation}
where $\beta$ denotes the trade-off parameter and $R\left(\mathbf{b}\right)$ is a regularization term on $\mathbf{b}$.
Note that here all historical information is captured in the first term, thus the regularization term only concerns the next portfolio, which is different from the EG algorithm.
One typical regularization is a L2-norm, that is, $R\left(\mathbf{b}\right)=\left\| \mathbf{b} \right\|^{2}$.

\citeN{AHK+06} proposed the \textit{Online Newton Step} (ONS), by solving the optimization problem~\eqref{eq:algorithms-follow_winner-ftrl} with L2-norm regularization via online convex optimization technique~\cite{Zinkevich03,HKK+06,Hazan06,HAK07}.
Similar to Newton method for offline optimization, the basic idea is to replace the $\log$ term via its second-order Taylor expansion at $\mathbf{b}_{t}$, and then solve the problem for closed-form update scheme.
Finally, ONS' update rule~\cite[Lemma 2]{AHK+06} is,
\begin{equation}\nonumber
\mathbf{b}_{1}=\left(\frac{1}{m}, \dots, \frac{1}{m}\right),\quad
\mathbf{b}_{t+1}=\Pi_{\Delta_{m}}^{\mathbf{A}_{t}}\left(\delta \mathbf{A}_{t}^{-1}\mathbf{p}_{t}\right),
\end{equation}
with
\begin{equation}\nonumber
\mathbf{A}_{t}=\sum_{\tau= 1}^{t}\left(\frac{\mathbf{x}_{\tau}\mathbf{x}_{\tau}^{\top}}{\left(\mathbf{b}_{\tau}\cdot\mathbf{x}_{\tau}\right)^{2}}\right) +\mathbf{I}_{m},\quad
\mathbf{p}_{t}=\left(1+\frac{1}{\beta}\right)\sum_{\tau=1}^{t} \frac{\mathbf{x}_{\tau}}{\mathbf{b}_{\tau}\cdot \mathbf{x}_{\tau}},
\end{equation}
where $\beta$ is the trade-off parameter,  $\delta $ is a scale term, and $\Pi_{\Delta_{m}}^{\mathbf{A}_{t}}\left(\cdot\right)$ is an exact projection to the simplex domain.

ONS iteratively updates the first and second order information and the portfolio with a time cost of $O\left(m^{3}\right)$, which is irrelevant to the number of historical instances $t$.
The authors also proved ONS's regret bound~\cite[Theorem 1]{AHK+06} of $O\left(m^{1.5}\log (mn)\right)$, which is worse than Cover's UP and Dirichlet(1/2) weighted UP.

While FTRL or even the Follow-the-Winner category mainly focuses on the worst-case investing,~\citeANP{HK09}\citeyear{HK09,HK12} linked the worst-case model with  practically widely used average-case investing, that is, the Geometric Brownian Motion (GBM) model~\cite{Bachelier00,Osborne59,Cootner64}, which is a probabilistic model of stock returns.
The authors also designed an investment strategy that is universal in the worst-case and is capable of exploiting the GBM model.
Their algorithm, or so-called \textit{Exp-Concave-FTL}, follows a slightly different form of optimization problem~\eqref{eq:algorithms-follow_winner-ftrl} with L2-norm regularization,
\begin{equation}\nonumber
\mathbf{b}_{t+1} = \mathop{\arg \max}_{\mathbf{b}\in \Delta_{m}}
  \sum_{\tau = 1}^{t}\log\left(\mathbf{b}\cdot
      \mathbf{x}_{\tau} \right)-\frac{1}{2}\left\| \mathbf{b}
  \right\|^{2}.
\end{equation}
Similar to ONS, the optimization problem can be efficiently solved via online convex optimization technique.
The authors further analyzed its regret bound and linked it with the GBM model.
Linking the GBM model, the regret round~\cite[Theorem~1.1 and Corollary~1.2]{HK12} is ${O}\left(m\log \left(Q+m\right)\right)$, where $Q$ denotes the quadratic variability, calculated as $n-1$ times the sample variance of the sequence of price relative vectors.
Since $Q$ is typically much smaller than $n$, the regret bound significantly improves the ${O}\left(m\log n\right)$ bound.

Besides the improved regret bound, the authors also discussed the relationship of their algorithm's performance to  trading frequency. The authors asserted that increasing the trading frequency would
decrease the variance of the minimum variance CRP, that is, the more
frequently they trade, the more likely the payoff will be close to the
expected value. On the other hand, the regret stays the same even if
they trade more. Consequently, it is expected to see improved
performance of such algorithm as the trading frequency increases~\cite{AHK+06}.

\citeN{DB11} further extended the FTRL approach to a generalized meta-learning algorithm, i.e., \textit{Online Newton Update} (ONU), which guarantees that the overall performance is no worse than any convex combination of its underlying experts.

\subsubsection{Aggregating-type Algorithms}
\label{sec:algorithms-follow_winner-aa}

Though BCRP is the optimal strategy for an i.i.d. market, the i.i.d. assumption is controversial in real markets, so the optimal portfolio may not belong to CRP or fixed fraction portfolio.
Some algorithms have been designed to track a different set of experts.
The algorithms in this category share similar idea to the Meta-Learning Algorithms in Section~\ref{sec:algorithms-meta_algorithms}.
However, here the base experts are of a special class, that is, individual expert that invests fully on a single stock, while in general Meta-Learning Algorithms often apply to more complex experts from multiple classes.

\citeN{VW98} applied the \textit{Aggregating Algorithm} (AA)~\cite{Vovk90,Vovk97,Vovk99,Vovk01} to the online portfolio selection task, of which Cover's UP is a special case.
The general setting for AA is to define a countable or finite set of base experts and sequentially allocate the resource among multiple base experts in order to achieve a good performance that is no worse than any fixed combination of underlying experts.
While its general form is shown in Section~\ref{sec:algorithms-meta_algorithms-aa}, its portfolio update formula~\cite[Algorithm 1]{VW98} for online portfolio selection is
\begin{equation}\nonumber
\mathbf{b}_{t+1}=\frac{\int_{\Delta_{m}}\mathbf{b}\prod_{i=1}^{t-1}\left(\mathbf{b}\cdot \mathbf{x}_{t}\right)^{\eta}P_{0}\left(d\mathbf{b}\right)}{\int_{\Delta_{m}}\prod_{i=1}^{t-1}\left(\mathbf{b}\cdot \mathbf{x}_{t}\right)^{\eta}P_{0}\left(d\mathbf{b}\right)},
\end{equation}
where $P_{0}\left(d\mathbf{b}\right)$ denotes the prior weights of the experts.
As a special case, Cover's Universal Portfolios corresponds to AA with uniform prior distribution and $\eta=1$.

%Several further applications have been proposed for the AA algorithm.
\citeN{Singer97} proposed \textit{Switching Portfolios} (SP) to track a changing market, in which the stock's behaviors may change frequently.
Rather than the CRP class, SP decides a set of basic strategies, for example, the pure strategy that invests all wealth in one asset, and chooses a prior distribution over the set of strategies.
Based on the actual return of each strategy and the prior distribution, SP is able to select a portfolio for each period.
Upon this procedure, the author proposed two algorithms, both of which assumes that the duration of using a basic strategy follows Geometric distribution with a parameter of $\gamma$, which can be fixed or varied in time.
With fixed $\gamma$, the first version of SP has explicit update formula~\cite[Eq.~(6)]{Singer97},
\begin{equation}\nonumber
\mathbf{b}_{t+1}=\left(1-\gamma-\frac{\gamma}{m-1}\right)\mathbf{b}_{t} + \frac{\gamma}{m-1}.
\end{equation}
With a varying $\gamma$, SP has no explicit update.
The author also adopted the algorithm for transaction costs.
Theoretically, the author further gave the lower bound of SP's logarithmic wealth with respect to any underlying switching regime in hindsight~\cite[Theorem 2]{Singer97}.
Empirical evaluation on Cover's 2-stock pairs shows that SP can outperform UP, EG, and BCRP, in most cases.

%Several further applications have been proposed for the AA algorithm.
%\citeN{Singer97} proposed \textit{Switching Portfolios} (SP), which is a regime-based trading strategy, that is, SP switches among a set of strategies corresponding to different regimes.
%At the beginning of each period, SP combines all strategies with the prior distribution to construct a portfolio.
%The author proposed two switching portfolio strategies, both of which assume that the duration of using one base strategy is geometrically distributed.
%While the first strategy assumes a fixed distribution parameter, the second assumes the distribution of the parameter is dynamically changing with respect to the duration.
%%The author further discussed the cases with transaction costs.
%Theoretically, the author further gave the lower bound of the logarithmic total wealth achieved with respect to the best of the switching regimes.
%Empirical results show that SP can outperform UP, EG and BCRP.

\citeN{LS08} proposed the \textit{Gaussian Random Walk} (GRW) strategy, which switches among the base experts according to Gaussian distribution.
\citeN{KS07} extended SP to \textit{piecewise} fixed fraction strategies, which partitions the periods into different segments and transits among these segments.
The authors proofed the piecewise universality of their algorithm, which can achieve the performance of the \textit{optimal} piecewise fixed fraction strategy.
\citeN{KS08} extended~\citeN{KS07} to the cases of transaction costs.
\citeANP{KS09}\citeyear{KS09,KS10} further generalized~\citeN{KS07} to sequential decision problem.
\citeN{KSB08} proposed another piece wise universal portfolio selection strategy via context trees and~\citeN{KSB11} generalized to sequential decision problem via tree weighting.

The most interesting thing is that switching portfolios adopts the
notion of regime switching~\cite{Hamilton94,Hamilton08}, which is different from the underlying assumption of universal portfolio selection methods and seems to be more plausible than an i.i.d. market.
The regime switching is also applied to some state-of-the-art trading strategies~\cite{HARDY01,MRS09}.
However, this approach suffers from its distribution assumption, because Geometric and Gaussian distributions do not seem to fit the market well~\cite{Mandelbrot63,Cont01}.
This leads to other potential distributions that can better model the markets.

\subsection{Follow-the-Loser Approaches}
\label{sec:algorithms-follow_loser}

The underlying assumption for the optimality of BCRP strategy is that market is i.i.d., which however does not always hold for the real-world data and thus often results in inferior empirical performance, as observed in various previous literatures. Instead of tracking the winners, the \textit{Follow-the-Loser} approach is often characterized by transferring the wealth from winners to losers.
The underlying assumption underlying this approach is \textit{mean reversion}~\cite{BT85,PS88,LM90}, which means that the good (poor)-performing assets will perform poor (good) in the following periods.

To better understand the mean reversion principle, let us further analyze the behaviors of CRP in the Example~\ref{example:bcrp}~\cite{LZH+12}.

\begin{example}[Synthetic market by~\citeN{CG86}]
%Assume the same synthetic market price relative sequence in previous example.
%Assume a two-asset market with cash and one volatile asset and the price relative sequence is $\mathbf{x}_{1}^{n}=\left\{ \left(1, 2\right), \left(1, \frac{1}{2}\right), \left(1, 2\right), \dots  \right\}$.
%Let us consider the BAH with uniform initial portfolio $\mathbf{b}_{1}=\left( \frac{1}{2}, \frac{1}{2} \right)$ and the CRP with uniform portfolio $\mathbf{b}=\left( \frac{1}{2}, \frac{1}{2}\right)$.
%Clearly, since no asset grows in the long run, the final wealth of BAH equals the uniform weighted summation of two assets, which is roughly $1$ in the long run.
%On the other side, according to the analysis of Example~\ref{example:cg86}, the final cumulative wealth of CRP is roughly $\frac{9}{8}^{\frac{n}{2}}$.
As illustrated in Example~\ref{example:bcrp}, uniform CRP grows exponentially on the synthetic market. Now we analyze its portfolio update behaviors, which follows mean reversion, as shown in Table~\ref{tab:algorithms-follow_loser-crp}.

Suppose the initial CRP portfolio is $\left(\frac{1}{2},\frac{1}{2}\right)$ and at the end of the $1^{\mathrm{st}}$ period, the closing price adjusted portfolio holding becomes $\left(\frac{1}{3}, \frac{2}{3}\right)$ and corresponding cumulative wealth increases by a factor of $\frac{3}{2}$.
At the beginning of the $2^{\mathrm{nd}}$  period, CRP manager rebalances the portfolio to initial uniform portfolio by transferring the wealth from good-performing stock (B) to poor-performing stock (A), which actually follows the mean reversion principle.
Then its cumulative wealth changes by a factor of $\frac{3}{4}$ and the portfolio holding at the end of the $2^{\mathrm{nd}}$ period becomes $\left(\frac{2}{3}, \frac{1}{3}\right)$.
At the beginning of the $3^{\mathrm{rd}}$ period, the wealth transfer with the mean reversion idea continues.

In a word, CRP implicitly assumes that if one stock performs poor (good), it tends to perform good (poor) in the subsequent period, and thus transfers the weights from good-performing stocks to poor-performing stocks.

%Though the BAH strategy gains nothing, BCRP can achieve a growth rate of $\frac{9}{8}$ per two trading periods, which implicitly assumes that if one stock performs poor, it tends to perform good in the subsequent trading period.

\begin{table*}[htbp]
    \tbl{Example to illustrate the mean reversion trading idea. \label{tab:algorithms-follow_loser-crp}}{
    \begin{tabular}{|c|c|c|c|c|c|}
        \hline
        $\#$ Period & Price Relative (A,B) & CRP & CRP Return & Portfolio Holdings &  Notes \\
        \hline
        1 & $\left(1, 2\right)$ & $\left(\frac{1}{2}, \frac{1}{2}\right)$ &
        $\frac{3}{2}$ & $\left(\frac{1}{3}, \frac{2}{3}\right)$  &
        $B\longrightarrow A$ \\
        2 & $\left(1, \frac{1}{2}\right)$ & $\left(\frac{1}{2}, \frac{1}{2}\right)$ &
        $\frac{3}{4}$ &  $\left(\frac{2}{3}, \frac{1}{3}\right)$   &
        $A\longrightarrow B$ \\
        3 & $\left(1, 2\right)$ & $\left(\frac{1}{2}, \frac{1}{2}\right)$ &
        $\frac{3}{2}$ & $\left(\frac{1}{3}, \frac{2}{3}\right)$  &
        $B\longrightarrow A$ \\
        $\vdots$ &$\vdots$ & $\vdots$ & $\vdots$ & $\vdots$ & $\vdots$ \\
        \hline
    \end{tabular}}
\end{table*}
\end{example}

\subsubsection{Anti Correlation}
\label{sec:algorithms-follow_loser-anticor}

\citeANP{BEG03}\citeyear{BEG03,BEG04} proposed a Follow-the-Loser portfolio
strategy named \textit{Anti Correlation} (Anticor) strategy.
Rather than no distributional assumption like Cover's UP, Anticor strategy assumes that market follows the mean reversion principle.
To exploit the mean reversion property, it statistically makes bet on the consistency of positive lagged cross-correlation and negative auto-correlation.

To obtain a portfolio for the $t+1^{st}$ period, Anticor adopts logarithmic price relatives~\cite{Hull08} in two specific market windows, that is, $\mathbf{y}_{1}=\log\left(\mathbf{x}_{t-2w+1}^{t-w}\right)$ and $\mathbf{y}_{2}=\log\left(\mathbf{x}_{t-w+1}^{t}\right)$.
It then calculates the cross-correlation matrix between
 $\mathbf{y}_{1}$ and $\mathbf{y}_{2}$,
\begin{equation}\nonumber
\begin{split}
M_{cov}\left(i, j\right) &= \frac{1}{w-1}\left(\mathbf{y}_{1, i}-\bar{y}_{1}\right)^{\top}\left(\mathbf{y}_{2, j}-\bar{y}_{2}\right)\\
M_{cor}\left(i, j\right) &=\begin{cases}
 \frac{M_{cov}\left(i, j\right)} {\sigma_{1}\left(i\right)*\sigma_{2}\left(j\right)} & \sigma_{1}\left(i\right), \sigma_{2}\left(j\right)\neq 0\\
 0 & \mbox{otherwise}
 \end{cases}
\end{split}
\end{equation}
Then according to the cross-correlation matrix, Anticor algorithm transfers the wealth according to the mean reversion trading idea, that is, moves the proportions from the stocks increased more to the stocks increased less, and the corresponding amounts are adjusted according to the cross-correlation matrix.
In particular, if asset $i$ increases more than asset $j$ and their sequences in the window are positively correlated, Anticor claims a transfer from asset $i$ to $j$ with the amount equals the cross correlation value ($M_{cor}\left(i, j\right)$) minus their negative auto correlation values ($\min\left\{ 0, M_{cor}\left(i, i\right) \right\}$ and $\min\left\{ 0, M_{cor}\left(j, j\right)\right\}$).
These transfer claims are finally normalized to keep the portfolio in the simplex domain.

Since its mean reversion nature, it is difficult to obtain a useful bound such as the universal regret bound.
Although heuristic and has no theoretical guarantee, Anticor empirically outperforms all other strategies at the time.
On the other hand, though Anticor algorithm obtains good performance outperforming all algorithms at the time, its heuristic nature can not fully exploit the mean reversion property.
Thus, exploiting the property using systematic learning algorithms is highly desired.

\subsubsection{Passive Aggressive Mean Reversion}
\label{sec:algorithms-follow_loser-pamr}

\citeN{LZH+12} proposed \textit{Passive Aggressive Mean Reversion} (PAMR) strategy, which exploits the mean reversion property with the Passive Aggressive (PA) online learning~\cite{SCD+03,CDK+06}.

The main idea of  PAMR is to design a loss function in order to reflect the mean reversion property, that is, if the expected return based on last price relative is larger than a threshold, the loss will linearly increase; otherwise, the loss is zero.
In particular, the authors defined the $\epsilon$-insensitive loss function for the $t^{th}$ period as,
\begin{equation}\nonumber
\ell_{\epsilon}\left(\mathbf{b}; \mathbf{x}_{t}\right)=\begin{cases}
0 & \mathbf{b}\cdot \mathbf{x}_{t}\leq \epsilon\\
\mathbf{b}\cdot \mathbf{x}_{t}-\epsilon & \mbox{otherwise}\\
\end{cases},
\end{equation}
where $0\leq \epsilon \leq 1$ is a sensitivity parameter to control the mean reversion threshold.
Based on the loss function, PAMR passively maintains last portfolio if the loss is zero, otherwise it aggressively approaches a new portfolio that can force the loss zero.
In summary, PAMR obtains the next portfolio via the following optimization problem,
\begin{equation}\label{eq:algorithms-follow_loser-pamr}
\mathbf{b}_{t+1} = \mathop{\arg\min}_{\mathbf{b}\in \Delta_{m}}\frac{1}{2}\left\| \mathbf{b}-\mathbf{b}_{t}\right\|^{2}\quad\mbox{s.t.}\quad\ell_{\epsilon}\left(\mathbf{b}; \mathbf{x}_{t}\right)=0.
\end{equation}
%To solve the situations with noisy price relatives in real-world financial markets, the authors also formulated two variants by introducing some non-negative slack variables into the optimization.
Solving the optimization problem~\eqref{eq:algorithms-follow_loser-pamr}, PAMR has a clean closed form update scheme~\cite[Proposition 1]{LZH+12},
\begin{equation}\nonumber
\mathbf{b}_{t+1}=\mathbf{b}_{t}-\tau_{t}\left(\mathbf{x}_{t}-\bar{x}_{t}\mathbf{1} \right),\quad\tau_{t}=\max\left\{0, \frac{\mathbf{b}_{t}\cdot \mathbf{x}_{t}-\epsilon}{\left\| \mathbf{x}_{t}-\bar{x}_{t}\mathbf{1}\right\|^{2}} \right\}.
\end{equation}
Since the authors ignored the non-negativity constraint of the portfolio in the derivation, they also added a simplex projection step~\cite{DSS+08}.
The closed form update scheme clearly reflects the mean reversion trading idea by transferring the wealth from the good performing stocks to the poor performing stocks.
It also coincides with the general form~\cite[Eq.~(1)]{LM90} of return-based contrarian strategies~\cite{CK98,Lo08}, except an adaptive multiplier $\tau_{t}$.
Besides the optimization problem~\eqref{eq:algorithms-follow_loser-pamr}, the authors also proposed two variants to avoid  noise price relatives, by introducing some non-negative slack variables into optimization, which is similar to  soft margin support vector machines.

Similar to Anticor algorithm, due to PAMR's mean reversion nature, it is hard to obtain a meaningful theoretical regret bound.
Nevertheless, PAMR achieves significant performance beating all  algorithms at the time and shows its robustness along with the parameters.
It also enjoys linear update time and runs extremely fast in the back tests, which show its practicability to large scale real world application.

The underlying idea is to exploit the single period mean reversion, which is empirically verified by its evaluations on several real market datasets.
However, PAMR suffers from drawbacks in risk management since it suffers significant performance degradation if the underlying single period mean reversion fails to exist.
Such drawback is clearly indicated by its performance in DJIA dataset~\cite{BEG03,BEG04,LZH+12}.

\subsubsection{Confidence Weighted Mean Reversion}
\label{sec:algorithms-follow_loser-cwmr}

\citeN{LHZ+11b} proposed \textit{Confidence Weighted Mean Reversion} (CWMR) algorithm to further exploit the second order portfolio information, which refers to the variance of portfolio weights (not price or price relative), following the mean reversion trading idea via Confidence Weighted (CW) online learning~\cite{DCP08,CDP08,CDK09,DKC10}.

The basic idea of CWMR is to model the portfolio vector as a multivariate Gaussian distribution with mean $\mathbf{\mu}\in \mathbb{R}^{m}$ and the diagonal covariance matrix $\mathbf{\Sigma}\in\mathbb{R}^{m\times m}$, which has nonzero diagonal elements $\sigma^{2}$ and zero for off-diagonal elements.
While the mean represents the knowledge for the portfolio, the diagonal covariance matrix term stands for the confidence we have in the corresponding portfolio mean.
Then CWMR sequentially updates the mean and covariance matrix of the Gaussian distribution and draws portfolios from the distribution at the beginning of a period.
In particular, the authors define $\mathbf{b}_{t}\in \mathcal{N}\left(\mathbf{\mu}_{t}, \mathbf{\Sigma}_{t}\right)$ and update the distribution parameters according to the similar idea of PA learning, that is, CWMR keeps the next distribution close to the last distribution in terms of   Kullback-Leibler divergence  if the probability of a portfolio return lower than $\epsilon$ is higher than a specified threshold.
In summary, the optimization problem to be solved is,
\begin{equation}\nonumber
\begin{split}
    \left(\mathbf{\mu}_{t+1}, \mathbf{\Sigma}_{t+1}\right)=
    \mathop{\arg\min}_{\mathbf{\mu}\in \Delta_{m}, \mathbf{\Sigma}}\quad
    &\mbox{D}_{\scriptscriptstyle \mbox{KL}}\left(\mathcal{N}\left(\mathbf{\mu},
    \mathbf{\Sigma}\right)\| \mathcal{N}\left(\mathbf{\mu}_{t},
    \mathbf{\Sigma}_{t}\right)\right)\\
    \mbox{s.t.}\quad&\mbox{Pr}\left[ \mathbf{\mu}\cdot
    \mathbf{x}_{t}
    \leq \epsilon\right]\geq \theta.\\
%    &\mathbf{\mu}\in \Delta_{m}.
\end{split}
\end{equation}
To solve the optimization,~\citeN{LHZ+13} transformed the optimization problem using two techniques.
One transformed optimization problem~\cite[Eq.~(3)]{LHZ+13} is,
\begin{equation}\nonumber%\label{eq:cwmr_var_formulation}
  \begin{split}
    \left(\mathbf{\mu}_{t+1}, \mathbf{\Sigma}_{t+1}\right)=\arg\min\quad
    &\frac{1}{2}\left(\log\left(\frac{\mbox{det}\mathbf{\Sigma}_{t}}{\mbox{det}\mathbf{\Sigma}}\right)
    +\mbox{Tr}\left(\mathbf{\Sigma}_{t}^{-1}\mathbf{\Sigma}\right)
    +\left(\mathbf{\mu}_{t}-\mathbf{\mu}\right)^{\top}\mathbf{\Sigma}_{t}^{-1}
      \left(\mathbf{\mu}_{t}-\mathbf{\mu}\right)\right)  \\
    \mbox{s. t.}\quad&\epsilon-\mathbf{\mu}^{\top}\mathbf{x}_{t}
    \geq\phi\mathbf{x}_{t}^{\top}\mathbf{\Sigma}\mathbf{x}_{t}\\
    & \mathbf{\mu}^{\top} \mathbf{1}=1,\quad \mathbf{\mu}\succeq 0.
  \end{split}
\end{equation}
%\begin{equation}\nonumber
%    \begin{split}
%      \left(\mathbf{\mu}_{t+1}, \mathbf{\Sigma}_{t+1}\right)=
%      \mathop{\arg\min}_{\mathbf{\mu}, \mathbf{\Sigma}}\quad
%      &\frac{1}{2}\left(\log\left(\frac{\mbox{det}\mathbf{\Sigma}_{t}}{\mbox{det}\mathbf{\Sigma}}\right)+
%        \mbox{Tr}\left(\mathbf{\Sigma}_{t}^{-1}\mathbf{\Sigma}\right)\right)
%      +\frac{1}{2}\left(\left(\mathbf{\mu}_{t}-\mathbf{\mu}\right)^{\top}\mathbf{\Sigma}_{t}^{-1}
%        \left(\mathbf{\mu}_{t}-\mathbf{\mu}\right)\right)
%      \\
%      \mbox{s. t.}\quad&\epsilon-\log\left(\mathbf{\mu}\cdot
%        \mathbf{x}_{t}\right) \geq \phi
%      \mathbf{x}_{t}^{\top}\mathbf{\Sigma}\mathbf{x}_{t}\\
%      & \mathbf{\mu}\cdot \mathbf{1}=1,\; \mathbf{\mu}\succeq 0.
%    \end{split}
%\end{equation}
%Note that the $\log$ in the constraint is manually substituted to utilize the $\log$ utility.
Solving the above optimization, one can obtain the closed form update scheme~\cite[Proposition~4.1]{LHZ+13} as,
  \begin{equation}\nonumber
    \begin{split}
      &\mathbf{\mu}_{t+1}=\mathbf{\mu}_{t}-\lambda_{t+1}\mathbf{\Sigma}_{t}
      \left(\mathbf{x}_{t}-\bar{x}_{t}\mathbf{1}\right),
      \quad \mathbf{\Sigma}_{t+1}^{-1}=\mathbf{\Sigma}_{t}^{-1}+2
      \lambda_{t+1}\phi\mathbf{x}_{t}\mathbf{x}_{t}^{\top},\\
    \end{split}
  \end{equation}
where $\lambda_{t+1}$ corresponds to the Lagrangian multiplier
calculated by Eq.~(11) in~\citeN{LHZ+13} and
$\bar{x}_{t}=\frac{\mathbf{1}^{\top}\mathbf{\Sigma}_{t}\mathbf{x}_{t}}{\mathbf{1}^{\top}\mathbf{\Sigma}_{t}\mathbf{1}}$
denotes the confidence weighted price relative average.
Clearly, the update scheme reflects the mean reversion trading idea and can exploit both the first and second order information of a portfolio vector.

Similar to Anticor and PAMR, CWMR's mean reversion nature makes it hard to obtain a meaningful theoretical regret bound for the algorithm.
Empirical performance show that the algorithm can outperform the state-of-the-art, including PAMR, which only exploits the first order information of a portfolio vector.
However, CWMR also exploits the single period mean reversion, which suffers the same drawback as PAMR.

\subsubsection{Online Moving Average Reversion}
\label{sec:algorithms-follow_loser-olmar}

Observing that PAMR and CWMR implicitly assume \textit{single-period} mean reversion, which causes one failure case on real dataset~\cite[DJIA dataset]{LZH+12},~\citeN{LH12} defined a \textit{multiple-period} mean reversion named \textit{Moving Average Reversion}, and proposed \textit{OnLine Moving Average Reversion} (OLMAR) to exploit the multiple-period mean reversion.

The basic intuition of OLMAR is the observation that PAMR and CWMR implicitly predicts next prices as last price, that is, $\hat{\mathbf{p}}_{t+1}=\mathbf{p}_{t-1}$, where $\mathbf{p}$ denotes the price vector corresponding the related $\mathbf{x}$.
Such extreme single period prediction may cause some drawbacks that caused the failure of certain cases in~\cite{LZH+12}.
Instead, the authors proposed a multiple period mean reversion, which explicitly predicts the next price vector as the moving average within a window.
They adopted simple moving average, which is calculated as $\mbox{MA}_{t} = \frac{1}{w}\sum_{i=t-w+1}^{t}\mathbf{p}_{i}$.
Then, the corresponding next price relative~\cite[Eq.~(1)]{LH12} equals,
\begin{equation}\label{eq:olmar_x}
\begin{split}
\hat{\mathbf{x}}_{t+1}\left(w\right) = \frac{MA_{t}\left(w\right)}{\mathbf{p}_{t}}
=&\frac{1}{w}\left(1+\frac{1}{\mathbf{x}_{t}}+\dots+\frac{1}{\bigodot_{i=0}^{w-2}\mathbf{x}_{t-i}} \right),
\end{split}
\end{equation}
where $w$ is the window size and $\bigodot$ denotes element-wise product.

Then, they adopted Passive Aggressive online learning~\cite{CDK+06} to learn a portfolio, which is similar to PAMR.
\begin{equation}\nonumber
\begin{split}
\mathbf{b}_{t+1} = \mathop{\arg\min}_{\mathbf{b}\in\Delta_{m}}\quad&\frac{1}{2}\left\|\mathbf{b}-\mathbf{b}_{t} \right\|^{2}\;
\mbox{s.t.}\; \mathbf{b}\cdot \hat{\mathbf{x}}_{t+1} \geq \epsilon
\end{split}.
\end{equation}
Different from PAMR, its formulation follows the basic intuitive of investment, that is, to achieve a good performance based on the prediction.
Solving the algorithm is similar to PAMR, and we ignore its solution.
At the time, OLMAR achieves the best results among all existing algorithms~\cite{LH12}, especially on certain datasets that failed PAMR and CWMR.

\subsubsection{Robust Median Reversion}
\label{sec:algorithms-follow_loser-rmr}

As existing mean reversion algorithms do not consider noises and outliers in the data, they often suffer from estimation errors, which lead to non-optimal portfolios and subsequent poor performance in practice.
To handle the noises and outliers,~\citeN{HZL+13} proposed to exploit mean reversion via robust L$_1$-median estimator, and designed a novel portfolio selection strategy called \textit{Robust Median Reversion} (RMR).

The basic idea of RMR is to explicitly estimate next price vector via robust L$_1$-median estimator at the end of $t^{th}$ period, that is, $\hat{\mathbf{p}}_{t+1}=L_1med_{t+1}\left(w\right)=\mathbf{\mu}_{t+1}$, where $w$ is a window size, and $\mathbf{\mu}$ is calculated by solving a optimization~\cite[Fermat-Weber problem]{Weber1909},
\begin{equation}\nonumber
    \mathbf{\mu}_{t+1} = \mathop{\arg\min}_{\mathbf{\mu}}\sum_{i=0}^{w-1}\left\|\mathbf{p}_{t-i}-\mathbf{\mu}  \right\|.
\end{equation}
In other words, L$_1$-median is the point with minimal sum of Euclidean distance to the $k$ given price vectors.
The solution to the optimization problem is unique~\cite{Weiszfeld1937} if the data points are not collinear.
Therefore, the expected price relative with L$_1$-median estimator becomes,
\begin{equation}\label{eq:rmr}
\hat{x}_{t+1}\left(w\right) = \frac{L_1med_{t+1}\left(w\right)}{\mathbf{p}_{t}}=\frac{\mathbf{\mu}_{t+1}}{\mathbf{p}_{t}}.
\end{equation}
Then RMR follows the similar portfolio optimization method as OLMAR~\cite{LH12} to learn an optimal portfolio.
Empirically, RMR outperforms the state-of-the-art on most datasets.

\subsection{Pattern-Matching based Approaches}
\label{sec:algorithms-pattern_based}

Besides the two categories of Follow-the-Winner/Loser, another type of strategies may utilize both winners and losers, which is based on pattern matching.
This category mainly covers nonparametric sequential investment strategies, which guarantee universal consistency, i.e., the corresponding trading rules are of growth optimal for any stationary and ergodic market process.
Note that different from the optimality of BCRP for the i.i.d. market, which motivates the Follow-The-Winner approaches, Pattern-Matching based approaches consider the non i.i.d. market and maximize the conditional expectation of \textit{log}-return given past observations (cf.~\citeN{AC88}).
For non i.i.d. market there is a big difference between the optimal growth rate and the growth rate of BCRP.
For example, for NYSE data sets during 1962-2006 the Average Annual Yield (AAY) of BCRP is about 20\%, while the strategies in this category have AAY more than 30\% (cf.~\citeN[Chapter 2]{GOW12}).
Grounded on nonparametric prediction~\cite{GS03}, this category consists of several pattern-matching based investment strategies~\cite{GLU06,GUV07,GUW08,LHG11a}.
Moreover, some techniques are also applied to the sequential prediction problem~\cite{BBG+10}.
%\citeN{GOW12} collect some related papers in this category.

Now let us describe the main idea of the Pattern-Matching based approaches~\cite{GLU06}, which consists of two steps, that is, the Sample Selection step and Portfolio Optimization step~\footnote{Here we only introduce the key idea.  All algorithms in this category consist of an additional aggregation step, which is a special case of Meta-Learning Algorithms in Section~\ref{sec:algorithms-meta_algorithms}.}.
The first step, Sample Selection step, selects an index set $C$ of similar historical price relatives, whose corresponding price relatives will be used to predict the next price relative.
After locating the similarity set, each sample price relative $\mathbf{x}_{i}, i\in C$ is assigned with a probability $P_{i}, i \in C$.
Existing methods often set the probabilities to uniform probability $P_{i}=\frac{1}{\left|C \right|}$, where $\left| \cdot \right|$ denotes the cardinality of a set.
Besides uniform probability, it is possible to design a different probability setting.
The second step, Portfolio Optimization step, is to learn an optimal portfolio based on the similarity set obtained in the first step, that is,
\begin{equation}\nonumber
    \begin{split}
      \mathbf{b}_{t+1}=\mathop{\arg\max}_{\mathbf{b} \in
        \Delta_{m}}\quad&U \left(\mathbf{b}; C \right),
    \end{split}
\end{equation}
where $U\left(\mathbf{b}; C\right)$ is a specified utility function of $\mathbf{b}$ based on $C$.
One particular utility function is the $\log$ utility, i.e., $U\left(\mathbf{b}; C\right)=\sum_{i\in C}\log \mathbf{b}^{\top}\mathbf{x}_{i}$, which is usually the default utility.
In case of empty similarity set, a uniform portfolio is adopted as the optimal portfolio.

In the following sections, we concretize the Sample Selection step in Section~\ref{sec:algorithms-pattern_based-sample_selection} and the Portfolio Optimization step in Section~\ref{sec:algorithms-pattern_based-portfolio_optimization}.
We further combine the two steps in order to formulate specific online portfolio selection algorithms, in Section~\ref{sec:algorithms-pattern_based-combinations}.

\subsubsection{Sample Selection Techniques}
\label{sec:algorithms-pattern_based-sample_selection}

The general idea in this step is to select similar samples from  historical price relatives by comparing the preceding market windows of two price relatives.
Suppose we are going to locate the price relatives that are similar to next price relative $\mathbf{x}_{t+1}$.
The basic routine is to iterate all historic price relative vectors $\mathbf{x}_{i}, i = w+1, \dots, t$ and count $\mathbf{x}_{i}$ as  similar one, if the preceding market window $\mathbf{x}_{i-w}^{i-1}$ is similar to the latest market window $\mathbf{x}_{t-w+1}^{t}$.
The set $C$ is maintained to contain the indexes of similar price relatives.
Note that market window is a $w\times m$-matrix and the similarity between two market windows is often calculated on the concatenated $w\times m$-vector.
The Sample Selection procedure ($C\left(\mathbf{x}_{1}^{t}, w\right)$) is further illustrated in Algorithm~\ref{alg:algorithms-pattern_based-sample_selection}.

\begin{algorithm}[t]
    \SetAlgoLined
    \KwIn{$\mathbf{x}_{1}^{t}$: Historical market sequence; $w$: window size;}
    \KwOut{${C}$: Index set of similar price relatives.}
    \BlankLine
    Initialize  ${C} =\emptyset$\;
    \If{$t\leq w+1$}{
        return;
    }
    \For{$i=w+1, w+2, \ldots, t$}{
        \If{$\mathbf{x}_{i-w}^{i-1}$ is similar to $\mathbf{x}_{t-w+1}^{t}$}{
                $C  = C \cup \{i\}$\;
        }
    }
    \caption{Sample selection framework ($C\left(\mathbf{x}_{1}^{t}, w\right)$).}
    \label{alg:algorithms-pattern_based-sample_selection}
\end{algorithm}

Nonparametric \textit{histogram-based} sample selection~\cite{GS03} pre-defines a set of discretized partitions, and partitions both latest market window $\mathbf{x}_{t-w+1}^{t}$ and historical market window $\mathbf{x}_{i-w}^{i-1}, i = w+1, \dots, t$, and finally chooses price relative vectors whose $\mathbf{x}_{i-w}^{i-1}$ is in the same partition as $\mathbf{x}_{t-w+1}^{t}$.
In particular, given a partition $P_{\ell}=A_{j, \ell}, j=1, 2, \dots, d_{\ell}$ of $\mathbb{R}_{+}^{m}$  into $d_{\ell}$ disjoint sets and a corresponding discretization function $G_{\ell}\left(\mathbf{x} \right)=j$, {if} $\mathbf{x}\in A_{\ell, j} $, we can define the similarity set as,
\begin{equation}\nonumber
C_{H}\left(\mathbf{x}_{1}^{t}, w\right) = \left\{ w < i < t+1 :
  G_{\ell}\left(\mathbf{x}_{t-w+1}^{t}\right)=G_{\ell}\left(\mathbf{x}_{i-w}^{i-1}\right)\right\}.
\end{equation}
Note that $\ell$ is adopted to aggregate multiple experts.

Nonparametric \textit{kernel-based} sample selection~\cite{GLU06} identifies the similarity set
by comparing two market windows via Euclidean distance,
\begin{equation}\nonumber
C_{K}\left(\mathbf{x}_{1}^{t}, w\right) = \left\{ w < i < t+1 :
 \left\| \mathbf{x}_{t-w+1}^{t}-\mathbf{x}_{i-w}^{i-1}\right\| \leq \frac{c}{\ell}\right\},
\end{equation}
where $c$ and $\ell$ are the thresholds used to control the number of similar samples.
Note that the authors adopted two threshold parameters for theoretical analysis.

Nonparametric \textit{nearest neighbor-based} sample selection~\cite{GUW08} searches the price relatives whose preceding market windows are within the $\ell$
nearest neighbor of latest market window in terms of Euclidean
distance,
\begin{equation}\nonumber
C_{N}\left(\mathbf{x}_{1}^{t}, w\right) = \left\{ w < i < t+1: \mathbf{x}_{i-w}^{i-1} \mbox{ is among the } \ell \mbox{
  NNs of } \mathbf{x}_{t-w+1}^{t}\right\},
\end{equation}
where $\ell$ is a threshold parameter.

\textit{Correlation-driven}  nonparametric sample selection~\cite{LHG11a} identifies the linear similarity among two market windows via correlation coefficient,
\begin{equation}\nonumber
C_{C}\left(\mathbf{x}_{1}^{t}, w\right) = \left\{  w < i < t+1:
  \frac{cov\left(\mathbf{x}_{i-w}^{i-1},
      \mathbf{x}_{t-w+1}^{t}\right)}{std\left(\mathbf{x}_{i-w}^{i-1}
    \right)std\left(\mathbf{x}_{t-w+1}^{t} \right)}\geq \rho\right\},
\end{equation}
where $\rho$ is a pre-defined correlation coefficient threshold.

\subsubsection{Portfolio Optimization Techniques}
\label{sec:algorithms-pattern_based-portfolio_optimization}

The second step of the Pattern-Matching based Approaches is to construct an optimal portfolio based on the similar set $C$.
Two main approaches are the Kelly's capital growth theory and Markowitz's mean variance theory.
In the following we illustrate several techniques adopted in this approaches.

\citeN{GLU06} proposed to figure out a \textit{log-optimal} (Kelly) portfolio based on similar price relatives located in the first step, which is clearly following the Capital Growth Theory.
Given a similarity set, the log-optimal utility function is defined as,
\begin{equation}\nonumber
U_{L}\left(\mathbf{b}; C\left(\mathbf{x}_{1}^{t}\right)\right) = \mathbb{E}\left\{\log \mathbf{b}\cdot \mathbf{x} \Big| \mathbf{x}_{i}, i\in C\left(\mathbf{x}_{1}^{t}\right) \right\} = \sum_{i\in C\left(\mathbf{x}_{1}^{t}\right)}P_{i}\log \mathbf{b}\cdot \mathbf{x}_{i},
\end{equation}
where $P_{i}$ denotes the probability assigned to a similar price relative $\mathbf{x}_{i}, i\in C\left(\mathbf{x}_{1}^{t}\right)$.
\citeN{GLU06} assume a uniform probability among the similar samples, thus it is equivalent to the following utility function,
\begin{equation}\nonumber
U_{L}\left(\mathbf{b}; C\left(\mathbf{x}_{1}^{t}\right)\right) = \sum_{i\in C\left(\mathbf{x}_{1}^{t}\right)}\log \mathbf{b}\cdot \mathbf{x}_{i}.
\end{equation}

\citeN{GUV07} introduced \textit{semi-log-optimal} strategy, which approximates $\log$ in the log-optimal utility function aiming to release the computational issue, and~\citeN{Vajda06} presented theoretical analysis and proved its universal consistency.
The semi-log-optimal utility function is defined as,
\begin{equation}\nonumber
U_{S}\left(\mathbf{b}; C\left(\mathbf{x}_{1}^{t}\right)\right) = \mathbb{E}\left\{f
 \left(  \mathbf{b}\cdot \mathbf{x} \right) \Big| \mathbf{x}_{i}, i\in C\left(\mathbf{x}_{1}^{t}\right) \right\} = \sum_{i\in C\left(\mathbf{x}_{1}^{t}\right)} P_{i}f\left(\mathbf{b}\cdot \mathbf{x}_{i}\right),
\end{equation}
where $f\left(\cdot\right)$ is defined as the second order Taylor expansion of $\log z$ with respect to $z=1$, that is,
\begin{equation}\nonumber
f\left(z\right) = z - 1 - \frac{1}{2}\left(z-1\right)^{2}.
\end{equation}
\citeN{GUV07} assume a uniform probability among the similar samples, thus, equivalently,
\begin{equation}\nonumber
U_{S}\left(\mathbf{b}; C\left(\mathbf{x}_{1}^{t}\right)\right) =  \sum_{i\in C\left(\mathbf{x}_{1}^{t}\right)} f\left(\mathbf{b}\cdot \mathbf{x}_{i}\right).
\end{equation}

\citeN{OV07} proposed nonparametric \textit{Markowitz-type} strategy, which is a further generalization of the semi-log-optimal strategy.
The basic idea of the Markowitz-type strategy is to represent portfolio return using Markowitz's idea to trade off between portfolio mean and variance.
To be specific, the Markowitz-type utility function is defined as,
\begin{equation}\nonumber
\begin{split}
U_{M}\left(\mathbf{b}; C\left(\mathbf{x}_{1}^{t}\right)\right) =&\mathbb{E}\left\{ \mathbf{b}\cdot \mathbf{x}\Big|\mathbf{x}_{i}, i \in C\left(\mathbf{x}_{1}^{t}\right) \right\}-\lambda \mathrm{Var}\left\{ \mathbf{b}\cdot \mathbf{x}\Big|\mathbf{x}_{i}, i \in C\left(\mathbf{x}_{1}^{t}\right)  \right\}\\
=&\mathbb{E}\left\{ \mathbf{b}\cdot \mathbf{x}\Big|\mathbf{x}_{i}, i \in C\left(\mathbf{x}_{1}^{t}\right) \right\} -\lambda  \mathbb{E}\left\{ \left(\mathbf{b}\cdot \mathbf{x}\right)^{2}\Big| \mathbf{x}_{i}, i \in C\left(\mathbf{x}_{1}^{t}\right) \right\}\\
+&\lambda \left( \mathbb{E}\left\{ \mathbf{b}\cdot \mathbf{x}\Big|\mathbf{x}_{i}, i \in C\left(\mathbf{x}_{1}^{t}\right) \right\}\right)^{2},
\end{split}
\end{equation}
where $\lambda$ is a trade-off parameter.
In particular, a simple numerical transformation shows that semi-log-optimal portfolio is an instance of the \textit{log-optimal} utility function with a specified $\lambda$.

To solve the problem with transaction costs,~\citeN{GV08} propose a \textit{GV-type} utility function (Algorithm $2$ in~\citeN{GV08}, their Algorithm $1$ follows the same procedure as $\log$-optimal utility) by incorporating the transaction costs, as follows,
\begin{equation}\nonumber
U_{T}\left(\mathbf{b}; C\left(\mathbf{x}_{1}^{t}\right)\right) =   \mathbb{E}\left\{ \log \mathbf{b}\cdot\mathbf{x} +\log c\left(\mathbf{b}_{t}, \mathbf{b}, \mathbf{x}_{t}\right)\right\},
\end{equation}
where $c\left(\cdot\right)\in \left(0, 1 \right)$ is the transaction cost factor in Eq.~\eqref{eq:problem_setting-transaction_cost}, which represents the remaining proportion after transaction costs imposed by the market.
The details of the calculation of the factor are illustrated in Section~\ref{sec:problem_setting-transaction_cost}.
According to a uniform probability assumption of the similarity set, it is equivalent to calculate,
\begin{equation}\nonumber
U_{T}\left(\mathbf{b}; C\left(\mathbf{x}_{1}^{t}\right)\right) =   \sum_{i\in C\left(\mathbf{x}_{1}^{t}\right)}\left( \log \mathbf{b}\cdot\mathbf{x} +\log c\left(\mathbf{b}_{t}, \mathbf{b}, \mathbf{x}_{t}\right)\right).
\end{equation}

%\if 0
%Observing that the transaction cost incurs when the aiming portfolio deviates from last closing price adjusted price relative,~\citeN{LHG11a} propose regularized nonparametric learning (RNL) and its main objective is to constraint the deviation from last closing price adjusted portfolio via regularization.
%In particular, they adopt a \textit{regularized log-optimal} utility function, that is,
%\begin{equation}\nonumber
%U_{R}\left(\mathbf{b}, C\left(\mathbf{x}_{1}^{t}\right)\right) =  \mathbb{E}\left\{ \log \mathbf{b}\cdot\mathbf{x} +\lambda \left\|  \hat{\mathbf{b}}_{t}- \mathbf{b}\right\|^{p}\right\},
%\end{equation}
%where $\lambda$ is the trade off parameter and $p\geq 1$ is the p-norm used to constrain the deviation from last closing price relative adjusted portfolio and $\hat{\mathbf{b}}_{t}$ is defined as, \begin{equation}\nonumber
%\hat{b}_{t, i} = \frac{b_{t, i}x_{t, i}}{\mathbf{b}_{t}\cdot \mathbf{x}_{t}},\quad i = 1, \dots, m.
%\end{equation}
%Clearly, assuming uniform probability among the similar samples as~\citeN{LHG11a}, it is equivalent to calculate,
%\begin{equation}\nonumber
%U_{R}\left(\mathbf{b}, C\left(\mathbf{x}_{1}^{t}\right)\right) = \sum_{i\in C\left(\mathbf{x}_{1}^{t}\right)} \left( \log \mathbf{b}\cdot\mathbf{x} +\lambda \left\|  \hat{\mathbf{b}}_{t}- \mathbf{b}\right\|^{p}\right),
%\end{equation}
%\fi

In any of the above procedures, if the similarity set is non-empty, we can gain an optimal portfolio based on the similar price relatives and their probability.
In case of empty  set, we can choose either  uniform portfolio or  previous portfolio.

\subsubsection{Combinations}
\label{sec:algorithms-pattern_based-combinations}

%Above two steps, combining each of the first step and each  of the second step will give one complete on-line portfolio strategy.
In this section, let us combine the first step and the second step and describe the detail algorithms in the Pattern-Matching based approaches.
Table~\ref{tab:algorithms_pattern_combination} shows existing combinations, where ``---"  means that no algorithm is proposed to exploit the combination.

\begin{table*}[htbp]
    \tbl{Pattern-Matching based approaches: sample selection and portfolio optimization.
    \label{tab:algorithms_pattern_combination}}{
    \begin{tabular}{|l|c|c|c|c|}
        \hline
         & \multicolumn{4}{c|}{Sample Selection Techniques}\\
         \hline
         Portfolio Optimization & Histogram   & Kernel    & Nearest Neighbor  &  Correlation \\
         \hline
         Log-optimal & $\mathrm{B}^\mathrm{H}$: C$_{H}$ + U$_{L}$  & $\mathrm{B}^\mathrm{K}$: C$_{K}$ + U$_{L}$   & $\mathrm{B}^\mathrm{NN}$: C$_{N}$ + U$_{L}$  & CORN: C$_{C}$ + U$_{L}$ \\
         Semi log-optimal & --- & $\mathrm{B}^\mathrm{S}$: C$_{K}$ + U$_{S}$   & --- & --- \\
         Markowitz-type & ---  & $\mathrm{B}^\mathrm{M}$: C$_{K}$ + U$_{M}$  &  ---   & ---\\
         GV-type &  ---  & $\mathrm{B}^\mathrm{GV}$: C$_{K}$ + U$_{R}$ &  ---  & --- \\
         %Regularized log-optimal &     &  &    & C$_{C}$ + U$_{R}$  \\
        \hline
    \end{tabular}
    }
\end{table*}

One default utility function is the log-optimal function.% or the BCRP portfolio.
\citeN{GS03} introduced the \textit{nonparametric histogram-based log-optimal} investment strategy ($\mathrm{B}^\mathrm{H}$), which combines the histogram-based sample selection and log-optimal utility function and proved its universal consistency.
\citeN{GLU06} presented \textit{nonparametric kernel-based log-optimal} investment strategy ($\mathrm{B}^\mathrm{K}$), which combines the kernel-based sample selection and log-optimal utility function and proved its universal consistency.
\citeN{GUW08} proposed \textit{nonparametric nearest neighbor log-optimal} investment strategy ($\mathrm{B}^\mathrm{NN}$), which combines the  nearest neighbor  sample selection and log-optimal utility function and proofed its universal consistency.
\citeN{LHG11a} created \textit{correlation-driven nonparametric} learning approach (CORN) by combining the correlation driven sample selection and log-optimal utility function and showed its superior empirical performance over previous three combinations.
Besides the log-optimal utility function, several algorithms using different utility functions have been proposed.
\citeN{GUV07} proposed  \textit{nonparametric kernel-based semi-log-optimal} investment strategy ($\mathrm{B}^\mathrm{S}$) by combining the kernel-based sample selection and semi-log-optimal utility function to ease the computation of ($\mathrm{B}^\mathrm{K}$).
\citeN{OV07} proposed  \textit{nonparametric kernel-based Markowitz-type} investment strategy ($\mathrm{B}^\mathrm{M}$) by combining the kernel-based sample selection and Markowitz-type utility function to make trade-offs between the return (mean) and risk (variance) of expected portfolio return.
\citeN{GV08} proposed  \textit{nonparametric kernel-based GV-type} investment strategy ($\mathrm{B}^\mathrm{GV}$) by combining the kernel-based sample selection and GV-type utility function to construct  portfolios in case of transaction costs.
%\citeN{LHG11a} proposed  \textit{regularized nonparametric learning} framework to select the optimal portfolio via regularization to constrain the deviation from last closing price adjusted  portfolio.
%They implement the combinations of correlation-based sample selection and regularized-log-optimal utility function.
If the sequence of relative price vectors is a first order Markov chain with known distributions then their strategies are growth optimal.
For unknown distributions, \citeN{GW12} introduced empirical growth optimal algorithms.
\citeN{OU11} empirically analyzed the performance of log-optimal portfolio strategies with transaction costs.

Note that this section only introduces the key steps (or individual expert) in the Pattern-Matching based approaches,
while all the above algorithms also consist an additional aggregation step.
With different parameters ($w$, $\ell$, or $\rho$), one can get a family of portfolios, which are then aggregated into a final portfolio using exponential weighting~\cite{GLU06}.
Such rule is actually a meta algorithms, which we will introduce in the following section.

\subsection{Meta-Learning Algorithms}
\label{sec:algorithms-meta_algorithms}

Another category of research in the area of online portfolio selection is the \textit{Meta-Learning Algorithm} (MLA)~\cite{DB11}, which is closely related to expert learning~\cite{CL06} in the machine learning community. This is directly applicable to a ``Fund of funds", which delegates its portfolio assets  to other funds. In general, MLA assumes several base experts, either from same strategy class or different classes.
Each expert outputs a portfolio vector for the coming period, and MLA combines these portfolios to form a final portfolio, which is used for the next rebalancing.
MA algorithms are similar to algorithms in ``Follow-the-Winner" approaches, however, they are proposed to handle a broader class of experts, which CRP can serve as one special case.
On the one hand, MLA system can be used to smooth the final performance with respect to all underlying experts, especially when base experts are sensitive to certain environments/parameters.
On the other hand, combining a universal strategy and a heuristic algorithm, where it is not easy to obtain a theoretical bound, such as Anticor, etc., can provide the universal property to the whole MLA system.
Finally, MLA is able to combine all existing algorithms, thus providing a much broader area of application.

\subsubsection{Aggregating Algorithms}
\label{sec:algorithms-meta_algorithms-aa}

Besides the algorithms discussed in Section~\ref{sec:algorithms-follow_winner-aa}, the \textit{Aggregating Algorithm} (AA)~\cite{Vovk90,VW98} can also be generalized to include more sophisticated base experts.
Given a learning rate $\eta > 0$, a measurable set of experts $A$, and a prior distribution $P_{0}$ that assigns the initial weights to the experts, AA defines a loss function as $\ell\left(\mathbf{x}, \gamma\right)$ and $\gamma_{t}\left(\theta\right)$ as the action chosen by expert $\theta$ at time $t$.
At the beginning of each period $t=1, 2, \dots$, AA updates the experts' weights  as,
\begin{equation}\nonumber
P_{t+1}\left(A\right) = \int_{A}\beta^{\ell\left(\mathbf{x}_{t}, \gamma_{t}\left(\theta\right)\right)}P_{t}\left(d\theta\right),
\end{equation}
where  $\beta=e^{-\eta}$ and $P_{t}$ denotes the weights to the experts at time $t$.

\subsubsection{Fast Universalization}
\label{sec:algorithms-meta_algorithms-fast_universalization}

\citeANP{ADK02}\citeyear{ADK02,ADM04} proposed \textit{Fast Universalization} (FU), which extends Cover's Universal Portfolios~\cite{Cover91} from parameterized CRP class to a wide class of investment strategies, including trading strategies operating on a single stock and portfolio strategies allocating wealth among whole stock market.
FU's basic idea is to evenly split the wealth among a set of base experts, let these experts operate on their own, and finally pool their wealth.
%Let us denote a by $\mathcal{E}\left(\mathbf{w}\right)$ with parameter $w\in \mathbb{W}$,  where $\mathbb{W}$ is the experts' parameter space, and each expert outputs its portfolio.
%Formally, at the end of period $t$, FU decides next portfolio equals historical performance weighted average of all base experts, that is,
%\begin{equation}\nonumber
%\mathbf{b}_{t+1}=\frac{\int_{\mathbb{W}}\mathcal{E}_{t}\left(\mathbf{w}\right)S_{t}\left(\mathcal{E}\left(\mathbf{w}\right)\right)\textit{d}\mu\left(\mathbf{w}\right)} {\int_{\mathbb{W}}S_{t}\left(\mathcal{E}\left(\mathbf{w}\right)\right)\textit{d}\mu\left(\mathbf{w}\right)},
%\end{equation}
%where $\mu\left(\mathbf{w}\right)$ denotes the uniform measure over $\mathbb{W}$.
FU's update is similar to that of Cover's UP, and it also asymptotically achieves the wealth equal to an optimal fixed convex combination of base experts.
In cases that all experts are CRPs, FU is reduced to Cover's UP.

Besides the universalization in the continuous parameter space, various discrete buy and hold combinations have been adopted by various existing algorithms.
Rewritten in discrete form, its update can be straightforwardly obtained.
For example,~\citeANP{BEG03}\citeyear{BEG03,BEG04} adopted BAH strategy to combine Anticor experts with a finite number of window sizes.
\citeN{LZH+12} combined PAMR experts with a finite number of mean reversion thresholds.
Moreover, all Pattern-Matching based approaches in Section~\ref{sec:algorithms-pattern_based} used BAH to combine their underlying experts, also with a finite number of window sizes.

\subsubsection{Online Gradient \& Newton Updates}
\label{sec:algorithms-meta_algorithms-ogu_onu}

\citeN{DB11} proposed two meta optimization algorithms, named \textit{Online Gradient Update} (OGU) and \textit{Online Newton Update} (ONU), which are natural extensions of \textit{Exponential Gradient} (EG) and \textit{Online Newton Step} (ONS), respectively.
Since their updates and proofs are similar to their precedents, here we ignore their updates.
Theoretically, OGU and ONU can achieve the growth rate as the optimal convex combination of the underlying experts.
Particularly, if any base expert is universal, the final meta system enjoys the universal property.
Such a property is useful since a Meta-Learning Algorithm can incorporate a heuristic algorithm and a universal algorithm, whereby the final system enjoys the performance while keeping the universal property.

\subsubsection{Follow the Leading History}
\label{sec:algorithms-meta_algorithms-flh}

\citeN{HS09} proposed a \textit{Follow the Leading History} (FLH) algorithm for  changing environments.
FLH can incorporate various universal base experts, such as the ONS algorithm.
Its basic idea is to maintain a working set of finite experts, which are dynamically flowed in and dropped out according to their performance, and allocate the weights among the active working experts with a meta-learning algorithm, for example, the Herbster-Warmuth algorithm~\cite{HW98}.
Different from other meta-learning algorithms where experts operate from the same beginning, FLH adopts experts starting from different periods.
Theoretically, the FLH algorithm with universal methods is universal.
Empirically, FLH equipped with ONS can significantly outperform ONS.

\section{Connection with Capital Growth Theory}
\label{sec:cgt}

Most online portfolio selection algorithms introduced above can be interestingly connected to the Capital Growth Theory. In this section, we first introduce the Capital Growth Theory for portfolio selection, and
then connect the previous algorithms to the Capital Growth Theory in order to reveal their underlying trading principles.

\subsection{Capital Growth Theory for Portfolio Selection}
\label{sec:cgt-cgt}

Originally introduced in the context of gambling, Capital Growth Theory (CGT)\cite{HZ95} (also termed Kelly investment~\cite{Kelly56} or Growth Optimal Portfolio (GOP) \cite{GOW12}) can generally be adopted for online portfolio selection.
\citeN{Breiman61} generalized Kelly criterion to multiple investment opportunities. \citeN{Thorp71} and~\citeN{Hakansson71} focused on the theory of Kelly criterion or logarithmic utility for the portfolio selection problem. Now let us briefly introduce the theory for portfolio selection~\cite{Thorp71}.

The basic procedure of CGT for portfolio selection is to maximize the expected $\log$ return for each period.
It involves two related steps, that is, prediction and portfolio optimization.
For prediction step, CGT receives  the predicted distribution of price relative combinations $\hat{\mathbf{x}}_{t{+}1}=\left(\hat{x}_{t{+}1, 1}, \dots, \hat{x}_{t{+}1, m}\right)$, which can be obtained as follows.
For each investment $i$, one can predict a finite number of distinct values and corresponding probabilities.
Let the range of $\hat{x}_{t{+}1, i}$ be $\left\{ r_{i, 1}; \dots; r_{i, {N}_{i}} \right\}, i = 1, \dots, m$, and corresponding probability for each possible value $r_{i, j}$ be $p_{i, j}$.
Based on these predictions, one can estimate their joint vectors and corresponding joint probabilities.
In this way, there are in total $\prod_{i=1}^{m}N_{i}$ possible prediction combinations, each of which is in the  form of $\hat{\mathbf{x}}_{t{+}1}^{\left(k_{1}, k_{2}, \dots, k_{m}\right)}=\left[\hat{x}_{t{+}1, 1}=r_{1, k_{1}}\mbox{ and } \hat{x}_{t{+}1, 2}=r_{2, k_{2}}\mbox{ and }\dots\mbox{ and  }\hat{x}_{t{+}1, m}=r_{m, k_{m}}\right]$ with a probability of $p^{\left(k_{1}, k_{2}, \dots, k_{m}\right)}=\prod_{j = 1}^{m}p_{j, k_{j}}$.
Given these predictions, CGT tries to obtain an optimal portfolio that maximizes the expected $\log$ return,
\begin{equation}\nonumber
\begin{split}
\mathbb{E}\log S &=  \sum p^{\left(k_{1}, k_{2}, \dots, k_{m}\right)}\log \left(\mathbf{b}\cdot \hat{\mathbf{x}}_{t{+}1}^{\left(k_{1}, k_{2}, \dots, k_{m}\right)}\right)\\
&= \sum\left[p^{\left(k_{1}, k_{2}, \dots, k_{m}\right)}\log \left(b_{1}r_{1, k_{1}}+\dots+b_{m}r_{m, k_{m}}\right) \right],\\
\end{split}
\end{equation}
where the summation is over all $\prod_{i=1}^{m}N_{i}$ price relative combinations.
Obviously, maximizing the above equation is concave in $\mathbf{b}$ and thus can be efficiently solved via convex optimization~\cite{BV04}.

\subsection{Online Portfolio Selection and Capital Growth Theory}
\label{sec:cgt-connection}

\begin{table}
\tbl{Online portfolio selection and the Capital Growth Theory.\label{tab:cgt-explicit}}{
  \begin{tabular}{l|l|l|l}
    \hline
    Algorithms & $\hat{\mathbf{x}}_{t+1}$ & Prob. & Capital growth forms \\
    \hline
    BCRP$^{\ast}$ & ${\mathbf{x}_{i}}, i=1, \dots, n$ & $1/n$ & $\mathbf{b}_{t+1}=\mathop{\arg\max}_{\mathbf{b}\in\Delta_{m}}\frac{1}{n}\sum_{i=1}^{n}\log \mathbf{b}\cdot {\mathbf{x}_{i}}$  \\
    \hline
    EG & $\mathbf{x}_{t}$ & $100\%$ & $\mathbf{b}_{t+1}=\mathop{\arg\max}_{\mathbf{b}\in\Delta_{m}}\log \mathbf{b}\cdot {\mathbf{x}_{t}} - \lambda R\left(\mathbf{b}, \mathbf{b}_{t}\right)$  \\
    PAMR & $\frac{1}{\mathbf{x}_{t}}$ & $100\%$ & $\mathbf{b}_{t+1}=\mathop{\arg\min}_{\mathbf{b}\in\Delta_{m}}\mathbf{b}\cdot {\mathbf{x}_{t}} + \lambda R\left(\mathbf{b}, \mathbf{b}_{t}\right)$  \\
    CWMR & $\frac{1}{\mathbf{x}_{t}}$ & $100\%$ & $\mathbf{b}_{t+1}=\mathop{\arg\min}_{\mathbf{b}\in\Delta_{m}} P\left( \mathbf{b}\cdot {\mathbf{x}_{t}} \right) + \lambda R\left(\mathbf{b}, \mathbf{b}_{t}\right)$  \\
    OLMAR/RMR & Eq.~\eqref{eq:olmar_x}/Eq.~\eqref{eq:rmr} & $100\%$ & $\mathbf{b}_{t+1}=\mathop{\arg\max}_{\mathbf{b}\in\Delta_{m}}\mathbf{b}\cdot {\hat{\mathbf{x}}_{t+1}} - \lambda R\left(\mathbf{b}, \mathbf{b}_{t}\right)$  \\
%    RMR & Eq.~\eqref{eq:rmr}& $100\%$ & $\mathbf{b}_{t+1}=\mathop{\arg\max}_{\mathbf{b}\in\Delta_{m}}\mathbf{b}\cdot {\hat{\mathbf{x}}_{t+1}} - \lambda R\left(\mathbf{b}, \mathbf{b}_{t}\right)$  \\
    \hline
     $\mathrm{B}^{\mathrm{H}}/{B}^{\mathrm{K}}/\mathrm{B}^{\mathrm{NN}}$/CORN & ${\mathbf{x}_{i}}, i \in C_{t}$ & $1/\left| C_{t}\right| $ & $\mathbf{b}_{t+1}=\mathop{\arg\max}_{\mathbf{b}\in\Delta_{m}}\frac{1}{\left|C_{t}\right|}\sum_{i\in C_{t}}\log \mathbf{b}\cdot {\mathbf{x}_{i}}$  \\
     $\mathrm{B}^{\mathrm{GV}}$ & ${\mathbf{x}_{i}}, i \in C_{t}$ & $1/\left| C_{t}\right| $ & $\mathbf{b}_{t+1}=\mathop{\arg\max}_{\mathbf{b}\in\Delta_{m}}\frac{1}{\left|C_{t}\right|}\sum_{i\in C_{t}}\left(\log \mathbf{b}\cdot {\mathbf{x}_{i}}+\log c\left(\cdot\right)\right)$  \\
     \hline
     FTL & ${\mathbf{x}_{i}}, i=1, \dots, t$ & $1/t$ & $\mathbf{b}_{t+1}=\mathop{\arg\max}_{\mathbf{b}\in\Delta_{m}}\frac{1}{t}\sum_{i=1}^{t}\log \mathbf{b}\cdot {\mathbf{x}_{i}}$  \\
     FTRL & ${\mathbf{x}_{i}}, i=1, \dots, t$ & $1/t$ & $\mathbf{b}_{t+1}=\mathop{\arg\max}_{\mathbf{b}\in\Delta_{m}}\frac{1}{t}\sum_{i=1}^{t}\log \mathbf{b}\cdot {\mathbf{x}_{i}} - \lambda R\left(\mathbf{b}\right)$  \\
    \hline
  \end{tabular}}
\end{table}

Most existing online portfolio selection algorithms have close connection with the Capital Growth Theory (or Kelly criterion).
While the theory provides a theoretically guaranteed framework for asset allocation, online portfolio selection algorithms mainly connect to the theory from two different aspects.

%The first connection is to assume that the market is i.i.d., then the theory shows that the maximum performance is achieved by BCRP~\cite[Theorem 15.3.1]{CT91}, and online portfolio algorithms (mainly Follow the Winner approaches) try to maximize the exponential growth rate relative to rate of BCRP.
%The second connection is to directly adopt the Capital Growth Theory for online portfolio selection.
%Table~\ref{tab:cgt-explicit} illustrates the algorithms that can be explicitly written in the form of CGT.

The first connection is established between CGT and the universal portfolio selection scheme, which mainly includes several Follow-the-Winner approaches.
Kelly criterion aims to maximize the exponential growth rate of an investment scheme, while universal portfolio selection scheme tries to maximize the exponential growth rate relative to the rate of BCRP.
Although they have different objectives, they are somehow connected as the target of universal portfolio selection scheme is one special case of Kelly criterion.
\citeN[Theorem~15.3.1]{CT91} showed that if the market sequence (price relative vectors) is i.i.d. and then the maximum performance is achieved by an optimal constant rebalanced portfolios in hindsight, or the Best Constant Rebalanced Portfolio (BCRP).
We further rewrite the BCRP strategy in the form of Kelly criterion, as shown in the first row of Table~\ref{tab:cgt-explicit}.
Thereafter,~\citeN{Cover91} set BCRP as a target, and proposed the universal portfolio selection scheme.
Note that as introduced in Section~\ref{sec:algorithms-benchmarks-crp}, the gap between their cumulative exponential grow rates is termed \textit{regret}.
Such connection also coincides with competitive analysis in~\citeN{BEG00}.

In particular, the first four algorithms in the Follow-the-Winner category, i.e., Universal Portfolios, Exponential Gradient, Follow the Leader, and Follow the Regularized Leader, all release regret bounds whose daily average asymptotically approaches zero as trading period goes to infinity.
In other words, these algorithms can achieve the same exponential growth rate as BCRP, which is CGT optimal in an i.i.d. market.
While the Aggregating-type Algorithms extend online portfolio selection from the CRP class to other strategy classes, which may not be optimal relative to BCRP.
%
%Mainly pursued by the Follow-the-Winner approaches, the first connect's underlying assumption is that the market sequence (price relative vectors) is i.i.d. and then the maximum performance is achieved by a constant rebalanced portfolios, or the Best Constant Rebalanced Portfolio (BCRP)~\cite[Theorem~15.3.1]{CT91} in hindsight, which is consistent to the principle of Capital Growth Theory.
%We further rewrite the BCRP strategy in the form of Capital Growth Theory, as shown in the first row of Table~\ref{tab:cgt-explicit}.
%Thereafter, pioneered by~\citeN{Cover91}, Follow the Winner approaches view BCRP as a target and try to maximize the exponential growth rate relative to the rate of BCRP.
%The gap between the two algorithms is also termed \textit{regret} illustrated as Eq.~\eqref{eq:algorithms-benchmarks-regret}.
%An algorithm with asymptotically zero regret is further called a \textit{universal} portfolio selection strategy.

The second connection explicitly adopts the idea of Capital Growth Theory for online portfolio selection, as shown in Section~\ref{sec:cgt-cgt}.
For each period, one algorithm requires the predicted price relative combinations and their corresponding probabilities.
Without loss of generality, let us make portfolio decision for the $t+1^{st}$ period.
Table~\ref{tab:cgt-explicit} summarizes their rewritten formulations.
Note that some algorithms in the first connection (such as EG, ONS, etc.) can also be rewritten to this form, although their objectives are different from CGT.
We present their implicit market distributions, denoted by their values ($\hat{\mathbf{x}}_{t+1}$) and probabilities (Prob.), in the second and third columns, respectively.
We then rewrite all algorithms following the Capital Growth Theory, that is, to maximize the expected $\log$ return for the $t+1^{st}$ period, in the fourth column.
The regularization terms are denoted as $R\left(\mathbf{b}, \mathbf{b}_{t}\right)$, which preserves the information of last portfolio vector ($\mathbf{b}_{t}$), and $R\left(\mathbf{b}\right)$, which constrains the variability of a portfolio vector.
Based on the number of predictions, we can categorize most existing algorithms into three categories.

The first category, including EG/PAMR/CWMR/OLMAR/RMR, implicitly or explicitly predicts a single scenario with certainty, and tries to select an optimal portfolio.
Note that the capital growth forms of PAMR and CWMR are rewritten from their original forms, while keeping their essential ideas.
Moreover, PAMR, CWMR, and OLMAR all ignore the $\log$ utility function, as adding $\log$ utility function follows the same idea but causes the convexity issue.
Though such single prediction is risky, all these algorithms adopt regularization terms, such as $R\left(\mathbf{b}, \mathbf{b}_{t}\right)=\left\| \mathbf{b}-\mathbf{b}_{t} \right\|^{2}$, to ensure that next portfolio is not far from current one, which in deed reduces the risk.

%It seems that such prediction is risky, however, all these algorithms adopt regularization terms, such as $R\left(\mathbf{b}, \mathbf{b}_{t}\right)=\left\| \mathbf{b}-\mathbf{b}_{t} \right\|^{2}$, to keep the information of last portfolio such that next portfolio is not far from last one, which in deed reduces the risk.

The second category, including Pattern-Matching based approaches, predicts multiple scenarios that are deemed similar to next price relative vector.
In particular, it expects next price relative to be $\mathbf{x}_{i}, i \in C$ with a uniform probability of $\frac{1}{\left|C\right|}$, where $C$ denotes the similarity set.
Then, algorithms in this category try to maximize its expected $\log$ return in terms of the similarity set, which is consistent with the Capital Growth Theory and results in an optimal fixed fraction portfolio.
Note that several algorithms in the Pattern-Matching based approaches, including $\mathrm{B}^{\mathrm{S}}$, $\mathrm{B}^{\mathrm{M}}$, and $\mathrm{B}^{\mathrm{GV}}$, adopt different portfolio optimization approaches, which we do not count in here.

The third category, including FTL and FTRL, implicitly predicts next scenario as all historical price relatives.
In particular, it predicts that the next price relative vector equals $\mathbf{x}_{i}, i = 1, \dots, t$ with a uniform probability of $\frac{1}{t}$.
Based on such prediction, strategies in this category aim to maximize the expected $\log$ return, and subtract a regularization term for FTRL.
Note that different from the regularization terms in the first category, regularization term in this category, such as  $R\left(\mathbf{b}\right)=\left\| \mathbf{b}\right\|^{2}$, only controls the deviation of next portfolio.
This is due to the fact that the predictions already contain all available information.

Note that the two connections are not exclusive. For example, some universal portfolio selection algorithms (EG, ONS, etc.) show both connections. On the one hand, their formulations can be explicitly rewritten to Kelly's form. On the other hand, their motivations follow the first connection, which is validated by their theoretical results.

\subsection{Underlying Trading Principles}
\label{sec:cgt-principle}

\begin{table*}
\tbl{Online portfolio selection and their underlying trading principles.\label{tab:cgt-principle}}{
  \begin{tabular}{l|l|l|l|l}
    \hline
    Principles & Algorithms & $\hat{\mathbf{x}}_{t+1}$ & Prob. & $\mathbb{E}\left\{\hat{\mathbf{x}}_{t+1}\right\}$ \\
    \hline
    \multirow{2}{*}{Momentum} & EG & $\mathbf{x}_{t}$ & $100\%$  & $\mathbf{x}_{t}$ \\
    & FTL/FTRL & ${\mathbf{x}_{i}}, i=1, \dots, t$ & $1/t$  & $\frac{1}{t}\sum_{i=1}^{t}\mathbf{x}_{i}$ \\
    \hline
    \multirow{5}{*}{Mean reversion} &  CRP/UP/AA & n/a & n/a & n/a\\
    &  Anticor & n/a & n/a & n/a\\
    &  PAMR/CWMR & $\frac{1}{\mathbf{x}_{t}}$ & $100\%$ & $\frac{1}{\mathbf{x}_{t}}$\\
    & OLMAR & Eq.~\eqref{eq:olmar_x} & $100\%$ & Eq.~\eqref{eq:olmar_x} \\
    & RMR & Eq.~\eqref{eq:rmr} & $100\%$ & Eq.~\eqref{eq:rmr} \\
    \hline
    Mixed & $\mathrm{B}^{\mathrm{H}}/\mathrm{B}^{\mathrm{K}}/\mathrm{B}^{\mathrm{NN}}$/$\mathrm{B}^{\mathrm{GV}}$/CORN & ${\mathbf{x}_{i}}, i \in C_{t}$ & $1/\left| C_{t}\right| $ & $\frac{1}{\left| C_{t}\right|}\sum_{i\in C_{t}}\mathbf{x}_{i}$  \\
    \hline
  \end{tabular}}
\end{table*}

Besides the aspect of the Capital Growth Theory, most existing algorithms also follow certain trading ideas to implicitly or explicitly predict their next price relatives.
Table~\ref{tab:cgt-principle} summarizes their underlying trading ideas via three trading principles, that is, momentum, mean reversion, and others (for example, nonparametric prediction).
%Note that the expected vector ($\mathbb{E}\left\{\hat{\mathbf{x}}_{t+1}\right\}$) is an element-wise operation.

Momentum strategy~\cite{CJL96,Rouwenhorst98,MG99,LS00,GH04,CGH04} assumes winners (losers) will still be winners (losers) in the following period.
By observing algorithms' underlying prediction schemes, we can classify EG/FTL/FTRL as this category.
While EG assumes that next price relative vector will be the same as last one, FTL and FTRL assume that next price relative is expected to be the average of all historical price relative vectors.

In contrary, mean reversion strategy~\cite{BT85,BT87,PS88,Jegadeesh91,CW03} assumes that winners (losers) will be losers (winners) in the following period. Clearly, CRP and UP, Anticor, and PAMR/CWMR belong to this category. Here, note that UP is an expert combination of the CRP strategies, and we classify it by its implicit assumption on the underlying stocks.
If we observe from the  perspective of experts, UP transfers wealth from CRP losers to CRP winners, which is actually momentum. Moreover, PAMR and CWMR's expected price relative vector is implicitly the inverse of last vector, which is in the opposite of EG.

Other trading ideas, including the Pattern-Matching based approach, cannot be classified as the above two categories.
For example, for the Pattern-Matching approaches, their average of the price relatives in a similarity set may be regarded as either momentum or mean reversion.
Besides, the classification of AA depends on the type of underlying experts.
From experts' perspective, AA always transfers the wealth from loser experts to winner experts, which is momentum strategy.
From stocks' perspective, which is the assumption in Table~\ref{tab:cgt-principle}, the classification of AA coincides with that of its underlying experts.
That is, if the underlying experts are single stock strategy, which is momentum, then we view AA's trading idea as momentum.
On the other hand, if the underlying experts are CRP strategy, which follows the mean reversion principle, we regard AA's trading idea as mean reversion.

% --------------------------------------------------------------
\section{Challenges and Future Directions}
\label{sec:open_issues}

Online portfolio selection task is a special and important case of asset management.
Though existing algorithms perform well theoretically or empirically in back tests, researchers have encountered several challenges in designing the algorithms.
In this section, we focus on the two consecutive steps in the online portfolio selection, that is, prediction and portfolio optimization.
In particular, we illustrate open challenges in the prediction step in Section~\ref{sec:open_issues_prediction}, and list several other challenges in the portfolio optimization step in Section~\ref{sec:open_issues_portfolio_optimization}.
There are a lot of opportunities in this area and and it worths further exploring.

\subsection{Accurate Prediction by Advanced Techniques}

\label{sec:open_issues_prediction}

As we analyzed  in Section~\ref{sec:cgt-connection}, existing algorithms implicitly assume various prediction schemes.
While current assumptions can result in good performance, they are far from perfect.
Thus, the challenges for the prediction step are often related to the design of more subtle prediction schemes in order to produce more accurate predictions of the price relative distribution.

--- Searching patterns. In the Pattern-Matching based approach, in spite of many sample selection techniques introduced, efficiently recognizing patterns in the financial markets is often challenging.
Moreover, existing algorithms always assume uniform probability on the similar samples, while it is an challenge to assign appropriate probability, hoping to predict  more accurately.
Finally, existing algorithms only consider the similarity between two market windows with the same length and same interval, however, locating patterns with varying timing~\cite{RL81,SC90,Keogh02} is also attractive.

--- Utilizing stylized facts in returns. In econometrics, there exist a lot \textit{stylized facts}, which refer to consistent empirical findings that are accepted as truth.
One stylized fact is related to autocorrelations in various assets' returns\footnote{Econometrics community often adopts simple net return, which equals price relative minus one.} .
It is often observed that some stocks/indices show positive daily/weekly/monthly autocorrelations~\cite{Fama70a,LM99,LMS+02}, while some others have negative daily autocorrelations~\cite{LM88,LM90,Jegadeesh90}.
An open challenge is to predict future price relatives utilizing their autocorrelations.

--- Utilizing stylized facts in absolute/square returns. Another stylized fact~\cite{Taylor05} is that the autocorrelations in absolute and squared returns are much higher than those in simple returns.
Such fact indicates that there may be consistent nonlinear relationship within the time series, which various machine learning techniques may be used to boost the prediction accuracy.
However, in the current prediction step, such information is rarely exploited, thus constituting a challenge.

--- Utilizing calendar effects. It is well known that there exist some \textit{calendar effects}, such as January effect or turn-of-the-year effect~\cite{RJ76,HL87,MZ08a}, holiday effect~\cite{Fields34,BM98,DZ10}, etc.
No existing algorithm exploits such information, which can potentially provide better predictions.
Thus, another open challenge is to take advantage of these calendar effects in the prediction step.

--- Exploiting additional information. Although most existing prediction schemes focus solely on the price relative (or price), there exists other useful side information, such as volume, fundamental, and experts' opinions, etc. \citeN{CO96} presented a preliminary model to incorporate the information, which is however far from applicability. Thus, it is an open challenge to incorporate other sources of information in order to facilitate the prediction of next price relatives.

\subsection{Open Issues of Portfolio Optimization}
\label{sec:open_issues_portfolio_optimization}

Portfolio optimization is the subsequent step for online portfolio selection.
While the Capital Growth Theory is effective in maximizing final cumulative wealth, which is the aim of our task, it often incurs high risk~\cite{Thorp97}, which is sometimes unacceptable for an investor.
Incorporating the risk concern to online portfolio selection is another open issue, which is not taken into account in current target.

--- Incorporating risk.  Mean Variance theory~\cite{Markowitz52} adopts variance as a proxy to risk.
However, simply adding variance may not efficiently trade off between return and risk.
Thus, one challenge is to exploit an effective proxy to risk and efficiently trade off between them in the scenario of online portfolio selection.

--- Utilizing ``optimal \textit{f}". One recent advancement in money management is ``optimal \textit{f}"~\cite{Vince90,Vince92,Vince95,Vince07,Vince09}, which is proposed to handle the drawbacks of Kelly's theory.
Optimal \textit{f} can reduce the risk of Kelly's approach, however, it requires an additional estimation on drawdown~\cite{MA04}, which is also difficult.
Thus, this poses one challenge to explore the power of optimal \textit{f} and efficiently incorporate it to the area of online portfolio selection.

--- Loosening portfolio constraints. Current portfolios are generally constrained in a simplex domain, which means the portfolio is self-financed and no margin/shorting. Besides current long-only portfolio, there also exist long/short portfolios~\cite{JL93}, which allow short selling and margin.
\citeN{Cover91} proposed a proxy to evaluate an algorithm when margin is allowed, by adding additional margin components for all assets.
%\citeN{CO98} proposed one universal method to exploit the portfolio when margin and short are allowed.
%% Added %%
Moreover, the empirical results on NYSE data~\cite[Chapter 4]{GOW12} show that for online portfolio selection and for short selling there is no gain, however, for leverage the increase of the growth rate is spectacular.
%% Added %%
However, current methods are still in their infancy, and far from application.
Thus, the challenge is to develop effective algorithms when margin and shorting are allowed.

--- Extending transaction costs. To make an algorithm practical, one has to consider some practical issues, such as transaction costs, etc.
Though several online portfolio selection models with transaction costs~\cite{BK99,Iyengar05,GV08} have been proposed, they can not be explicitly conveyed in an algorithmic perspective, which are hard to understand.
One challenge is to extend current algorithms to the cases when transaction costs are taken into account.

--- Extending market liquidity. %Another consideration is market liquidity.
Although all published algorithms claim that in the back tests they choose blue chip stocks, which have the highest liquidity, it can not solve the concern of market liquidity.
Completely solving this problem may involve paper trading or real trading, which is difficult for the community.
Besides, no algorithm has ever considered this issue in its algorithm formulation.
The challenge here is to accurately model the market liquidity, and then design efficient algorithms.

% --------------------------------------------------------------
\section{Conclusions}
\label{sec:conclusion}

This article conducted a survey on the online portfolio selection problem, an interdisciplinary topic of machine learning and finance.
With the focus on algorithmic aspects, we began by formulating the task as a sequential decision learning problem, and further categorized the existing algorithms into five major groups: Benchmarks, Follow-the-Winner, Follow-the-Loser, Pattern-Matching based approaches, and Meta-Learning algorithms.
After presenting the surveys of individual algorithms, we further connected them to the Capital Growth Theory in order to better understand the essence of their underlying trading ideas. Finally, we outlined some open challenges for further investigations. We note that, although quite a few algorithms have been proposed in literature, many open research problems remain unsolved and deserve further exploration. We wish this survey article could facilitate researchers to understand the state-of-the-art in this area and potentially inspire their further study.

\begin{acks}
The authors would like to thank VIVEKANAND GOPALKRISHNAN for his comments on an early discussion of this work, and AMIR SANI for his careful proofreading.
\end{acks}

% --------------------------------------------------------------
\bibliographystyle{acmsmall}
\bibliography{reference}

%\received{December 2012}{December 2012}{December 2012}

\end{document}